\documentclass[useAMS,usenatbib,usegraphicx]{mn2e}
\usepackage{graphicx,verbatim,natbib,amsmath,amssymb}

\newcommand{\rmd}{\ensuremath{\mathrm{d}}}
\newcommand{\photoz}{photo-$z$}
\newcommand{\zphot}{\ensuremath{z_{\rm phot}}}
\newcommand{\specz}{spec-$z$}
\newcommand{\zspec}{\ensuremath{z_{\rm spec}}}
\newcommand{\zlens}{\ensuremath{z_{\rm lens}}}
\newcommand{\zsrc}{\ensuremath{z_{\rm src}}}
\newcommand{\photozsigma}{\ensuremath\sigma_{\Delta z/(1+z)}}
\newcommand{\Mr}{\ensuremath{^{0.3}\!M_r}}
\newcommand{\MLR}{\ensuremath{M_*/L}}

\title[Lensing with photometric redshifts]{Photometric redshift requirements for lens galaxies in galaxy-galaxy lensing analyses}
\author[Nakajima et al.]{R. Nakajima$^{1,2,3}$\thanks{\texttt{rnakajima@ewha.ac.kr}}, 
  R. Mandelbaum$^4$, U. Seljak$^{1,2,3,5}$, J.D. Cohn$^{1,2}$, R. Reyes$^4$, 
\newauthor
  R. Cool$^{4}$\thanks{Hubble Fellow and Carnegie-Princeton Fellow.}
  \\	
  $^1$Space Sciences Lab, Department of Physics and Department of Astronomy, University of California, Berkeley, CA  94720, USA
  \\
  $^2$Lawrence Berkeley National Lab, University of California, Berkeley, CA  94720, USA
  \\
  $^3$Institute of the Early Universe, Ewha Womans University, Seoul, Korea 
  \\
  $^4$Department of Astrophysical Sciences, Princeton University, Peyton Hall, Princeton, NJ 08544, USA 	
  \\	
  $^5$Institute for Theoretical Physics, University of Zurich, Zurich, Switzerland
}

\begin{document}

\date{\today}

\maketitle

\begin{abstract}

Weak gravitational lensing is a valuable probe of galaxy formation and cosmology.
Here we quantify the effects of using photometric redshifts (\photoz) in galaxy-galaxy 
lensing, for both sources and lenses, both for the immediate goal of
using galaxies with \photoz\ as lenses in the Sloan Digital Sky Survey
(SDSS) and as a demonstration
of methodology for large, upcoming weak lensing surveys that will by
necessity be dominated by lens samples with \photoz.  We calculate the bias in the lensing mass 
calibration as well as consequences for absolute magnitude 
(i.e., $k$-corrections) and stellar mass estimates, for a large sample
of SDSS Data Release 8 (DR8) galaxies. 
The redshifts are obtained with the template based \photoz\ code
ZEBRA on the SDSS DR8 $ugriz$ photometry.  We assemble and characterise 
the calibration samples ($\sim$9k spectroscopic redshifts from four surveys) to 
obtain photometric 
redshift errors and lensing biases corresponding to our full SDSS DR8 lens and 
source catalogues.
Our tests of the calibration sample also highlight the impact of 
observing conditions in the imaging survey when the spectroscopic calibration 
covers a small fraction of its footprint; 
atypical imaging conditions in calibration fields
can lead to incorrect conclusions regarding the \photoz\ of the full survey.

For the SDSS DR8 catalogue, we find 
$\photozsigma=0.096$ and $0.113$ for the lens and source catalogues,
with flux limits of $r=21$ and $r=21.8$, respectively.
The \photoz\ bias and scatter is a function of \photoz\ and template types, which we exploit
to apply \photoz\ quality cuts.  By using \photoz\ rather than spectroscopy for
lenses, dim blue galaxies and $L_*$ galaxies up to $z\sim 0.4$ can be
used as lenses, thus expanding into unexplored areas of parameter space.
We also explore the systematic uncertainty in the lensing signal
calibration when using source \photoz, and both lens and source \photoz; 
given the size of existing training samples, we can constrain the lensing signal calibration
(and therefore the normalization of the surface mass density) to within
2 and 4 per cent, respectively.

\end{abstract}

\begin{keywords}
gravitational lensing: weak --- methods: data analysis --- galaxies:
distances and redshifts --- galaxies: photometry --- cosmology: observations
\end{keywords}

\section{Introduction}

The current $\Lambda$CDM cosmological model is dominated by the unknown dark components
of the universe: dark matter and dark energy \citep[e.g.,][]{komatsu_etal:2011}.
Gravitational lensing, the deflection of light from distant 
source galaxies by intervening masses along the line-of-sight 
\citep[e.g.,][]{bartelmann_schneider:2001, refregier:2003},
has emerged as an enormously powerful astrophysical and cosmological probe.  
It is not only sensitive to dark energy through both cosmological 
distance measures and large-scale structure growth \citep{albrecht_etal:2006},
but is also sensitive to all forms of matter, including dark matter.
Measurements of the statistical distortion of galaxy shapes (or weak
gravitational lensing) due to mass along the line-of-sight have been
used in numerous studies to constrain the cosmological model, the
theory of gravity, and the connection between galaxies and dark matter
\citep[e.g., most recently,][]{mandelbaum_etal:2006, fu_etal:2008, reyes_etal:2010,
schrabback_etal:2010}.
As a result, many more surveys are planned for the next two decades 
with weak lensing as a major science driver:
KIDS\footnote{\texttt{http://www.astro-wise.org/projects/KIDS/}},
DES\footnote{\texttt{https://www.darkenergysurvey.org/}},
HSC\footnote{\texttt{http://oir.asiaa.sinica.edu.tw/hsc.php}},
Pan-STARRS\footnote{\texttt{http://pan-starrs.ifa.hawaii.edu/public/}},
LSST\footnote{\texttt{http://www.lsst.org/lsst}},
Euclid\footnote{\texttt{http://sci.esa.int/science-e/www/area/index.cfm?fareaid=102}},
and WFIRST\footnote{\texttt{http://wfirst.gsfc.nasa.gov/}}.

In order to fully access the information encoded in gravitational lensing,
redshift information is essential, as the conversion from distortions
(gravitational shear) to mass
depends on the lens and source redshifts via the critical surface
density, expressed as 
\begin{equation}\label{E:sigmacrit}
\Sigma_{c}=\frac{c^2}{4\pi G} \frac{D_S}{(1+z_L)^2 D_L D_{LS}}
\end{equation}
in comoving coordinates (in physical coordinates, the expression lacks
the factor of $(1+z_L)^{-2}$).  Here, $D_L$ and $D_S$ are angular diameter distances to the lens and
source, and $D_{LS}$ is the angular diameter distance between the lens
and source. 
Spectroscopic redshifts provide the best accuracy in determining $\Sigma_c$, but obtaining 
them for large, statistical samples of both lenses and sources is
prohibitively expensive in terms of observing time and instrumentation.
As a result, upcoming surveys will rely heavily upon less accurate 
photometric redshifts (\photoz) derived from multi-band imaging.
Since we require source samples at higher redshift, and as the data pushes into
the region of dim galaxies with poor photometry, the \photoz\ may worsen 
even more.   Consequently, it is of immediate interest to quantify
how limited redshift accuracy (due to the use of photometric redshifts) propagates into
lensing results.

Galaxy-galaxy lensing has been used in the past to quantify the
connection between lens galaxies or clusters and their dark matter (DM) halos, in particular
the total (average) mass profile around galaxies on $>20$kpc scales
and the DM halo occupation statistics \citep{2005ApJ...635...73H,2006MNRAS.371L..60H,mandelbaum_etal:2006}; can constrain the dark matter power spectrum when used in
combination with the galaxy
2-point correlation function
\citep{yoo_etal:2006,cacciato_etal:2009,baldauf_etal:2010,2011PhRvD..83b3008O}; and can
constrain the
theory of gravity when combined with clustering and redshift-space
distortions \citep{2007PhRvL..99n1302Z,reyes_etal:2010}.  
This paper addresses the use of photometric redshifts
for both sources and lenses in galaxy-galaxy lensing, in the context of
the Sloan Digital Sky Survey Data Release 8
\citep[and references therein; SDSS DR8 hereafter]{eisenstein_etal:prep}.
Previous galaxy-galaxy lensing studies with SDSS have been limited to lenses with 
spectroscopic redshifts \citep[e.g.,
][]{2004AJ....127.2544S,mandelbaum_etal:2006} or photo-$z$ 
with atypically high accuracy, such as those for Brightest Cluster
Galaxies \citep[e.g., ][]{sheldon_etal:2009}.  Galaxy-galaxy lensing
studies with other surveys have typically either involved an unusually
large number of passbands yielding exceptionally good \photoz\
\citep[e.g., ][]{kleinheinrich_etal:2006,leauthaud_etal:2011}, or have
had limited area coverage and therefore relatively poor statistics
compared with the SDSS studies \citep[e.g., ][]{parker_etal:2007}.  

Additional work must be done to allow galaxy-galaxy lensing to achieve
its full potential with large, upcoming imaging surveys, and to extend
to lens galaxy samples that lack spectra in SDSS.
Lens galaxy samples that lack spectra in SDSS tend to be smaller, dimmer 
galaxies, or galaxies at higher redshifts: it would be interesting to extend
galaxy-galaxy lensing studies to their DM halo masses and environments,
including mass and redshift dependence.  While studies with other
surveys have extended into these regimes, they have typically involved
deep but very narrow space-based data with significant cosmic variance
\citep{2006MNRAS.371L..60H,leauthaud_etal:2011} or wider but still relatively 
noisy ground-based survey data \citep{2005ApJ...635...73H}. 

In order to address galaxy-galaxy lensing based on SDSS DR8 \photoz, we 
have calculated a new set of photometric redshifts for the full
flux-limited galaxy sample with extinction-corrected model $r<21.8$.  
We applied the publicly available \photoz\ code 
Z\"{u}rich Extragalactic Bayesian Redshift Analyzer (ZEBRA\footnote{\tt
http://www.exp-astro.phys.ethz.ch/ZEBRA/}, \citealt{feldmann_etal:2006}) 
to the SDSS $ugriz$ photometry.  In this work, 
we demonstrate the \photoz\ accuracy by comparing against several
spectroscopic samples.  In the course of this process, we identify
several concerns related to the observing conditions in the imaging
survey in regions that overlap the calibration survey, which will be
generally applicable to any upcoming survey.

We define the criterion for a ``better'' \photoz\ as a \photoz\ 
that gives minimal scatter in the distribution of $\zspec-\zphot$, 
and hence the lowest scatter in the biases of any redshift-derived
quantities, such as the lensing signal or the absolute magnitude.
Note that we do not aim for the lowest bias in \photoz, according to
the philosophy that biases can be corrected with a representative
calibration sample, but rather the lowest {\em uncertainty} in the 
bias\footnote{Since the scatter defines the uncertainty, we need a 
representative calibration sample that is large enough to calibrate it.  
For deeper surveys for which such a large representative sample may not 
necessarily be available, the definition of 'ideal photometric redshifts' 
may differ, with a preference for those that show little structure in the 
\photoz\ error as a function of redshift, luminosity, or type.}.  
This criterion results in a low uncertainty in the actual science analysis.  
A brief comparison to other publicly available \photoz\ catalogues is presented.
Improvements in the characterisation of \photoz\ for the photometric galaxies can
provide a substantial boost in statistics for current studies of the large scale structure.
In addition,
working with the \photoz\ to the survey limiting magnitude with SDSS photometry will
provide an ideal test case for future large surveys that rely upon \photoz\ for 
all lensing calculations (which are, necessarily, dominated by
galaxies near the flux limit).

With the improved and extended SDSS \photoz's, we consider the
application to galaxy-galaxy lensing.
Photo-$z$ for weak lensing source galaxies in SDSS was
investigated by \citealt{mandelbaum_etal:2008}.  Here, we address some
additional nuances in the procedures from that work to define a fair
calibration sample and to use it to estimate the lensing calibration
bias.  We then extend that formalism to use  \photoz\ for {\em lenses}, 
thus increasing the possible sample of lens galaxies by a
factor of $\sim 40$ over the flux-limited MAIN spectroscopic sample ($r<17.77$) and 
colour-selected LRGs (luminous red galaxies; $r<19.5$).
With an eye to using \photoz\ for galaxy-galaxy lensing, 
we then test how the bias and scatter in the photometric redshifts for the SDSS photometric galaxy sample
affect various derived quantities (including the lensing signal
calibration, and the estimated luminosities and stellar masses) by direct
comparison to their spectroscopic redshift (\specz) counterparts.
While \cite{kleinheinrich_etal:2005} identify the need for lens
redshift information over simply using a lens redshift distribution,
our work is the first detailed demonstration of how to calibrate the
effect of the lens \photoz\
errors on other lensing observables.

The outline of the paper is as follows:
we describe the data and calibration subsets in Sec.~\ref{sec:data}, 
and test the imaging quality in the regions overlapping the
calibration samples for consistency with the average survey quality in Sec.~\ref{sec:test_calib}. 
This latter step is crucial for ensuring that measurements
using the calibration set are representative of the full SDSS DR8 sample of interest.
We justify our choice of \photoz\ method in Sec.~\ref{sec:photoz_method}.
The photometric redshift accuracy results (bias and scatter) are discussed in
Sec.~\ref{sec:photoz_results}.
From the photometry and the \photoz, we also derive estimated absolute luminosity and stellar mass,
described in Sec.~\ref{ssec:absmag_bias} and Sec.~\ref{ssec:stellar_mass}, respectively.
Sec.~\ref{sec:application} then describes the
resulting biases of derived quantities for galaxy-galaxy lensing applications.
The lensing signal calibration is discussed separately in Sec.~\ref{sec:calib_lens},
and conclusions are presented in Sec.~\ref{sec:conclusion}.  
We use a flat concordance WMAP7 \citep{komatsu_etal:2011} cosmology 
($\Omega_m=0.27$, $h=0.702$) to calculate luminosity distances $D_L$ 
and angular diameter distances $D_A$ from redshifts.

\section{Data}
\label{sec:data}

Here we describe the data sets used for this investigation:  
the SDSS photometric catalogue and the spectroscopic calibration sets.

\subsection{SDSS photometry}\label{ssec:sdssphotom}
The Sloan Digital Sky Survey (SDSS, \citealt{SDSS:2000}) is a shallow,
wide-field survey that imaged 14~555 deg$^2$ of the sky to
$r\sim 22$, and followed up roughly two million of the detected
objects spectroscopically
\citep{eisenstein_etal:2001,richards_etal:2002,strauss_etal:2002}.
The five-band ($ugriz$) SDSS imaging
\citep{fukugita_etal:1996,smith_etal:2002} was carried out by
drift-scanning the sky in photometric conditions
\citep{hogg_etal:2001,ivezic_etal:2004,padmanabhan_etal:2008} using a
specially-designed wide-field camera
\citep{gunn_etal:1998}.   All of
the data were processed by completely automated pipelines that detect
and measure photometric properties of objects, and astrometrically
calibrate the data \citep{lupton_etal:2001, pier_etal:2003,tucker_etal:2006}. 
The original goals of SDSS were completed with its seventh data release 
\citep[DR7,][]{sdssdr7:2009}.

Accurate galaxy colours are needed in order to compute reliable \photoz's.
We have chosen our SDSS galaxy sample, described below, from the most recent 
SDSS-III Data Release 8 \citep[DR8]{eisenstein_etal:prep}.
The galaxy colours obtained are based on the model magnitudes {\tt MODELMAG} 
\citep{stoughton_etal:2002}.  The 5-band fluxes are estimated
using a single galaxy model (the better fit of exponential or de Vaucouleurs)
based on $r$-band imaging, which is then convolved with the band-specific point-spread 
function (PSF) and allowed to vary only in its amplitude in order to
estimate the flux in each band.  This procedure leads to a consistent
definition of the magnitudes across all bands, despite the different PSFs.
This method is superior to PSF-matching because
it does not require convolutions of the data (convolutions lead to correlated
noise, making estimation of the flux uncertainties challenging).
A correction for galactic extinction was imposed using the dust maps from
\cite{schlegel_etal:1998} and the extinction-to-reddening ratios from
\cite{sdssedr:2002}; we only use regions with $r$-band
extinction $A_r<0.2$.  This extinction cut is standard for many extragalactic observations, but is also particularly
important here since dust extinction affects the relative magnitudes of the different bands,
and the magnitude corrections become less reliable for higher extinctions. 
The photometry was calibrated using ubercalibration procedure
\citep{padmanabhan_etal:2008}, to ensure consistent calibration across
the entire survey (within $1$ per cent), which is also important for \photoz\
uniformity. 
Minor corrections were applied to the $u$ and $z$ band to correct to AB magnitudes
($-0.04$ and $+0.02$, 
respectively)\footnote{\tt http://cas.sdss.org/dr7/en/help/docs/algorithm.asp}
for calculating the photometric redshifts.
We note that using a more recent estimate on the absolute calibration, 
$m_{\rm AB} - m_{\rm SDSS} = -0.036, 0.012, 0.010, 0.028, 0.040$ for
the $ugriz$ bands\footnote{\tt
  http://howdy.physics.nyu.edu/index.php/Kcorrect}, does not
significantly modify our results.

Requirements on the overall data quality for a region to be used are that the
ubercalibration procedure \citep{padmanabhan_etal:2008} must classify
the data as photometric (CALIB\_STATUS $=1$); the $r$-band PSF FWHM must be smaller than
1.8$^{\prime\prime}$; and various {\sc Photo} flags must indicate no major problems
with the object detection and PSF estimation\footnote{We require
  $0\le {\rm IMAGE\_STATUS}\le 4$, ${\rm PSP\_STATUS}=0$, ${\rm PHOTO\_STATUS}=0$.}.
In the acceptable regions, a total of 8720 square degrees, 
additional cuts were imposed when selecting galaxies; see
Appendix~\ref{S:galcut} for details.  In regions with
multiple observations, we first chose the observation with the best
seeing, and then imposed the magnitude cut.  


The photometric (lens) catalogue is a purely flux-limited catalogue of galaxies with 
\photoz\ information (this work). 
The full SDSS photometric catalogue has $r<22$, but to avoid the
galaxies with very noisy photometry near the flux limit (based on
previous findings, e.g. \citealt{kleinheinrich_etal:2005}, that accurate lens redshift information is
more important than source redshift information), we require $r<21$ for our lens sample.  
 Galaxy detection is also more stable under different observing
 conditions with this magnitude cut, as discussed in
 Sec.~\ref{ssec:obs_cond_effect}, resulting in a relatively uniform lens
 galaxy number density across the survey. 
There are 28~036~133 objects in the photometric lens catalogue of $r<21$
(decreased from 64~750~701
in the full $r<22$ catalogue).

The source catalogue, in contrast, goes to a depth of $r<21.8$ 
and contains additional information about the galaxy shapes (Reyes et~al\@. 2011 in~prep.).
In order to derive galaxy shapes without 
significant systematic error, the source catalogue has further quality cuts in order to ensure that
the galaxies are sufficiently resolved compared to the PSF.   It is an update of the 
\citet{mandelbaum_etal:2005} source catalogue, with several technical improvements and 
with additional area.  
As for the old catalogue, the galaxy shape measurements are obtained using the REGLENS
pipeline, including PSF correction done via re-Gaussianization
\citep{hirata_seljak:2003} and with cuts designed to avoid various
shear calibration biases.  These cuts include a requirement that the
galaxy be well-resolved in both the $r$ and $i$ bands, since the
lensing measurements use an average of the $r$ and $i$ band shapes.
There are 43~378~516 objects in the source catalogue that satisfy the resolution 
requirement (\photoz\ quality cuts will reduce this further
by $\sim10$ per cent, as described in Sec.~\ref{ssec:photozbytemplates}).

\subsection{Calibration data sets}
\label{ssec:spectroset}

Quantities derived from \photoz's, such as absolute magnitudes and 
the lensing signal calibration, will be biased due to \photoz\
error\footnote{As demonstrated in \cite{mandelbaum_etal:2008}, even \photoz\ that are
  unbiased on average will cause a bias in the lensing signal due to
  the non-negligible scatter.  This is a consequence of the nonlinear
  dependence of the lensing critical surface density on the lens and
  source redshifts.}.
We estimate these biases (and their uncertainty) by
measuring them on a representative subset of the SDSS catalogue
for which spectroscopic redshifts are available.  This calibration set
must satisfy the following criteria:
(1) Target selection is based on apparent magnitude only,
with no colour selection (the latter often is used in spectroscopic surveys 
to select objects of certain redshift ranges and/or types)---unless
two samples with complementary colour selection can be combined to
make an effectively flux-limited survey. 
(2) If targeted off of different photometry than the imaging survey,
then the limiting apparent magnitude must be somewhat deeper than the photometric survey, 
because increased flux uncertainty at the limiting magnitude will randomly scatter objects 
below and above the magnitude threshold.   (3) The spectroscopic sample should not have
any redshift failure modes that have a strong preference for
particular galaxy types (magnitude, colour, or redshift).  Note that
some of these criteria are not absolute, in the sense that the
analysis we describe could be adapted for calibration data sets that do
not satisfy them perfectly, but it would significantly complicate the
analysis and result in greater systematic error.

\begin{figure}
\includegraphics[width=84mm]{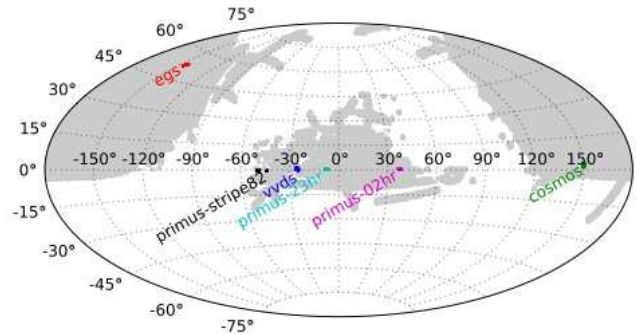}
\caption{
Footprints of the SDSS which satisfy our photometric quality criteria (8720 deg$^2$, shaded)
and spectroscopic calibration subsamples (labeled).
Each calibration subsample covers the full photometric depth of SDSS
(except PRIMUS-stripe82, see text), but only covers $\sim0.2$ per cent of the SDSS survey area.
}
\label{fig:calib_subsets} 
\end{figure}

There are a limited number of surveys with spectroscopic or
spectro-photometric redshift determinations
that meet these criteria.  
The rest of this section describes each survey that we use for this analysis, noting
in particular their survey targeting strategy, our quality cut criteria, and
the resulting number of calibration galaxies.
Figure~\ref{fig:calib_subsets} shows the footprints of SDSS and calibration subsamples;
Table~\ref{tab:calibcounts} summarises the number of galaxies from each survey that
have been matched to our SDSS catalogue. 
Further cuts are needed so that the calibration and full samples are representative of the 
same populations (Sec.~\ref{sec:test_calib}); there is an additional
\photoz\ quality cut (Sec.~\ref{ssec:photozbytemplates}).   These counts are shown
in Table~\ref{tab:calibcounts} in bold.
After the full set of cuts, the parent calibration sample has 9~631
sources and 8~592 lenses; the former constitutes a substantial
increase over \cite{mandelbaum_etal:2008}, which used 2~838 galaxies
from only two of the calibration samples described below.
While we have not addressed the spectroscopic failure 
rate on the calibration bias estimates, 
\citet[Sec.~5.5]{mandelbaum_etal:2008} studied the implications of a $\sim5$ per cent
failure rate and found minimal difference on the calibration bias.

\begin{table}
\begin{tabular}{@{}l @{}rr rr}
\hline
                  & Redshift     & Fraction & \multicolumn{2}{c}{Counts} \\
\cline{4-5} 
Survey            & completeness & secure & lens & source \\
\hline
EGS               & 93 \ \       & 99     & \ 639        & 1060    \\ 
                  &              &        & \ {\bf 593}  &  \vspace{1mm} {\bf 968} \\
zCOSMOS           & 85 \ \       & $>99.5$& \  775       & 1120    \\
pCOSMOS           & $>97$ \ \    & 97     & \ 2506       & 3644    \\
COSMOS (union)    &              &        & \ 3281       & 4764    \\
                  &              &        & \ {\bf 2959} & \vspace{1mm} {\bf 4293} \\
VVDS              & $55$ \ \     & 97     & \ 643        & 1132    \\
                  &              &        & \ {\bf 564}  &  \vspace{1mm} {\bf 527} \\
PRIMUS-02hr       & $93$ \ \ & $98$  & (1436)       & (2588)  \\
\ldots with DEEP2 &              &        & 1508         & 2908    \\
                  &              &        & \ {\bf 1357} & \vspace{1mm} {\bf 1646} \\
PRIMUS-23hr       & $90$ \ \ & $98$  & (1612)       & (3023)  \\
\ldots with DEEP2 &              &        & 1677         & 3307    \\
                  &              &        & \ {\bf 1547} & \vspace{1mm} {\bf 1737} \\
PRIMUS-stripe82   & 80   \ \     & $98$  & \ 856        &  -      \\
                  &              &        & \ {\bf 786}  &  \vspace{1mm} {\bf -}  \\
\hline
Total             &              &        & \ {\bf 8592} &  {\bf 9171} \\
\hline
\end{tabular}
%
%
\caption{A summary of the properties of our redshift calibration
  samples, including the redshift completeness (in per cent), the fraction of secure
  redshifts corresponding to that completeness (in per cent), and the original numbers of 
  lens and source galaxies available in each one.  
  The counts in bold are the actual number of objects used for calibration, after
seeing (Sec.~\ref{ssec:obs_correction}) and \photoz\ quality cuts
(Sec.~\ref{ssec:photozbytemplates}) have been applied.}
\label{tab:calibcounts}
\end{table}

\subsubsection{DEEP2 EGS}
The DEEP2 Galaxy Redshift Survey \citep{
davis_etal:2003, madgwick_etal:2003,coil_etal:2004,davis_etal:2005,davis_etal:2007}
is a spectroscopic survey using the DEep Imaging Multi-Object Spectrograph 
(DEIMOS, \citealt{faber_etal:2003}) on the Keck II Telescope 
at high spectral resolution ($R\sim5000$).
Of the four DEEP2 fields, we use \specz's from the Extended Groth Strip (EGS) field
at RA$\sim$14hr, in which the targets include galaxies of all colours with $R_{AB}<24.1$.   
For $R>21.5$, the target selection deviates slightly from a purely
flux-limited sample.  As a result, 
a small fraction ($\sim2$ per cent) of the SDSS-matched targets have been
down-weighted based on colour; faint objects with $21.5 < R_{AB} < 24.1$ 
and colours suggesting a redshift $z<0.75$ receive lower targeting priority.
However, this down-weighting was found to have minimal impact 
on the derived scatter and biases for the overall matched sample.
Due to saturation issues, no objects brighter than $R_{AB}=17.6$ were targeted;
these missing bright objects constitute a small fraction ($\lesssim3$ per cent) of both our lens 
and source samples.  
There is a known redshift incompleteness which we discuss later in Sec.~\ref{ssec:deficient_redshift_bins}.

DEEP2 provides redshift quality flags, of which we keep flags 3 and 4,  
which correspond to 95 and $>$99.5 per cent repeatability, respectively.
The good-quality ($q=3,4$) redshifts constitute 
$\sim85$ per cent of the SDSS-matched calibration set, and the fraction are similar
for both the lens and source sample.
Over half of the remaining objects have their redshifts confirmed by visual inspection, 
such that the SDSS-matched EGS sample 
is 99 per cent secure, with 92 per cent completeness for both the lens and source samples
(Table~\ref{tab:calibcounts}).
Of the DEEP2 EGS spectroscopic galaxies,  
639 and 1060 objects matched to the SDSS photometric catalogue ($r<21$) and source catalogue
($r<21.8$ with resolution cuts), respectively.  The latter number
differs from that in \cite{mandelbaum_etal:2008} because of the use of
a new reduction of the SDSS data to create a new source catalogue (see section~\ref{ssec:sdssphotom}).

\subsubsection{zCOSMOS}
The zCOSMOS-bright survey \citep{lilly_etal:2007, lilly_etal:2009} is a 
magnitude-limited spectroscopic survey on the Cosmological Evolution Survey (COSMOS) field
\citep{capak_etal:2007,scoville_etal:2007a,scoville_etal:2007b,taniguchi_etal:2007} 
with the VIsible Multi-Object Spectrograph (VIMOS, \citealt{lefevre_etal:2003}) on 
the 8m European Southern Observatory Very Large Telescope (ESO VLT)
at intermediate spectral resolution ($R\sim600$).
The selection is purely flux-limited at $15.0 < I_{AB} < 22.5$.
We choose objects with confidence class 3 and 4;
additionally, we include confidence class 9.5 objects, which have redshifts determined 
from a single emission line and which are consistent with the (30-band COSMOS)
photometric redshifts discussed in section~\ref{SSS:pcosmos};
these constitute $\sim$1 per cent of the SDSS-matched sample.  
The \photoz\ agreement is necessary to break the degeneracy between H$\alpha$ 
and [OII]3727, when the doublet appears as a single line due to line broadening.
Table~\ref{tab:calibcounts} lists the number of objects from this survey 
available for \photoz\ calibration.

Our matched zCOSMOS calibration sample is smaller than that from
\citet{mandelbaum_etal:2008} because 
$\sim 1/3$ of the SDSS imaging in the COSMOS region is classified
as non-photometric \citep{mandelbaum_etal:2010}, resulting in
abnormally poor \photoz.   Since the non-photometric regions are not
representative of our source catalogue (for which we impose a
photometricity cut), we eliminate the region of COSMOS for which the
SDSS data are not photometric from consideration in this analysis,
leaving 775 and 1120 lens
and source galaxies (before further cuts that will be 
described below).  An
additional consideration regarding the COSMOS calibration sample is
that the sky noise level in the SDSS imaging in the COSMOS region is significantly higher
than what is typical for the SDSS survey overall.   As we will show in Sec.~\ref{ssec:sdss_completeness}, the consequence is 
an atypical deficit of fainter galaxies in the COSMOS
calibration sample; we discuss how this deficit is handled in
Sec.~\ref{ssec:source_lensbias}.   

\subsubsection{COSMOS photometric redshifts (pCOSMOS)}
\label{SSS:pcosmos}

The first new calibration sample used in this work is the sample of
flux-limited non-zCOSMOS galaxies in the COSMOS region, with 
photometric redshifts \citep[pCOSMOS hereafter]{ilbert_etal:2009}.
Given the flux limit of the SDSS catalogue with respect to the COSMOS
observations, and the
accuracy demanded in the applications described in this paper, these
COSMOS \photoz\ are effectively the same as spectroscopic redshifts\footnote{That is,
we have verified that increasing the true redshift uncertainty by
up to a few per cent will not degrade our ability to estimate the bias
in derived redshift quantities, because the errors in the SDSS
\photoz\ will always dominate.}.

The pCOSMOS \photoz\ are 
obtained from a $\chi^2$ template-fitting method, 
{\tt Le Phare}\footnote{\tt http://www.oamp.fr/people/arnouts/LE\_PHARE.html}.
Highly accurate \photoz\ result from the use of deep photometry in 30
bands (primary bands are 
$u^*$, $B_J$, $V_J$, $g^+$, $r^+$, $i^+$, $z^+$, and $K$),
observed at various telescopes, primarily from Subaru and the
Canada-France-Hawaii Telescope (CFHT).
Comparing the COSMOS photo-z with the matching zCOSMOS redshifts, we estimate the scatter in
the pCOSMOS calibration subset to be $\sigma_{\Delta z/(1+z)}\sim0.009$, with an outlier rate 
of $\sim 1.2$ per cent for both the lens and the source samples (we find the bias to
be negligible).
%
COSMOS galaxies that have a zCOSMOS spectroscopic redshift have been removed, so
that the zCOSMOS and pCOSMOS catalogues are disjoint 
(though they trace similar large-scale structures).  
There are 2506 and 3644 pCOSMOS matches to the photometric 
lens and source catalogues, respectively.

\subsubsection{VVDS}
The next new calibration sample is the VIMOS VLT Deep Survey 
\citep[VVDS]{lefevre_etal:2005}, another spectroscopic survey 
using the VIMOS instrument on the ESO-VLT.  
We use the public spectroscopic catalogue in the VVDS-F22 field, part of the
VVDS-wide survey. This survey uses a lower resolution ($R\sim230$ 
versus $600$) but similar exposure time 
($3000$ versus $3600$ sec) compared to the zCOSMOS-bright survey.
The VVDS-wide target selection is purely magnitude-limited ($17.5<I_{AB}<22.5$) 
with no colour selection.
We select those galaxies with spectroscopic quality flags 3 and 4, with $\sim$96 and $\sim$99
per cent reliability, respectively.  When combined, these 
constitute only $\sim$55 per cent of the SDSS-matched sample,
suggesting a relatively low redshift success rate for this sample.
Additionally, galaxies with qualities 13, 14 (broad emission line) or  23, 24 (serendipitous observations) 
are included; however these constitute a small fraction (3 per cent) of the SDSS-matched sample.   
In total, we use 643 lens and 1132 source galaxies with VVDS
spectroscopic redshifts for SDSS \photoz\ calibration, see
Table~\ref{tab:calibcounts}. 
Because of concerns about the high spectroscopic redshift failure
rate, we have (1) checked the colour and magnitude distributions of the
VVDS-matched sample (Section~\ref{ssec:spectro_completeness}) and (2) done all of the calculations of biases due
to \photoz\ error 
separately for individual spectroscopic subsamples.  We find
(Section~\ref{ssec:source_lensbias}) no 
systematic tendency for the VVDS results to differ with any of the
others, suggesting that the redshift failure rate has not excessively
biased the calibration sample properties.  

\subsubsection{PRIMUS}
\label{ssec:primus}
The final additions to our calibration sample come from the PRIsm MUlti-object Survey 
\citep[PRIMUS;][Cool et~al\@. 2011 in~prep.]{coil_etal:prep}, a newly-completed survey 
that obtained spectro-photometric redshifts for $\sim120~000$ galaxies to $i=23$ over an area of 9.1 deg$^2$.  
On the Inamori Magellan Areal Camera and Spectrograph (IMACS) on the Magellan 
I Telescope, PRIMUS uses a low-dispersion prism with $R\sim40$ to efficiently 
survey wide areas and achieve a redshift accuracy of $\sigma_{
z/(1+z)}\lesssim 0.005$ out to $z=1.2$.
From the PRIMUS low-resolution spectra,
the general shape of the spectrum, in addition to emission or absorption lines, 
is used to infer the redshift (Cool et~al\@. 2011 in~prep.). 
PRIMUS is generally a flux-limited survey (no colour selection);
however there are a few relevant exceptions to this rule that we will
discuss shortly.
The target selection in most target fields used full and sparse sampling for 
$i<22.5$ and $22.5<i<23.5$, respectively, although the limiting
magnitudes and reference band for these categories vary depending on
the target field. 

We included data from 3 fields, which we will denote PRIMUS-02hr, PRIMUS-23hr, and 
PRIMUS-stripe82.  
The PRIMUS-02hr and 23hr fields overlap the DEEP2 02hr and 23hr spectroscopic survey fields
(these DEEP2 fields are $BRI$ colour-selected to include almost
exclusively $z>0.7$ galaxies).
The magnitude limits in these fields for full and sparse sampling were $R<22.8$ and $22.8<R<23.5$, respectively.  
The primary targeting strategy in these PRIMUS fields was to target the complement of the
DEEP2 spectroscopic survey;
hence the union of the PRIMUS and DEEP2 surveys 
in the 02hr and 23hr fields is essentially a flux-limited sample \citep{coil_etal:prep}.
We have verified that the relative weighting of DEEP2 to PRIMUS,
which accounts for the targeting efficiency and survey area differences,
is very near unity in the SDSS-matched sample in both fields.
A fraction of the PRIMUS targets also have high-quality redshifts in
DEEP2; for these, 
the spectroscopic redshifts from DEEP2 were used.
The supplements from the DEEP2 survey with quality flag $>=3$ constitute 
$\sim$5 and $\sim$10 per cent of photometric lens ($r<21$) and source ($r<21.8$) catalogues,
respectively.
We limit our PRIMUS redshifts to those having quality flag of 3 or 4.
The corresponding reliabilities are estimated at 93 and 97 per cent, respectively.

Like the other two fields, the PRIMUS-stripe82 calibration field
overlaps the SDSS stripe-82 region, but in this case the PRIMUS
targeting was carried out from single-epoch SDSS imaging to $r<22$ or
$i<21$ with no colour selection.  The fact that this targeting was
carried out using a different (earlier) reduction of the SDSS imaging,
in one particular observing run, results in targeting incompleteness 
at the faint end, with respect to the DR8 photometry.  This issue, which is caused by the scatter in the
derived magnitudes between different reductions, is apparent in
Figure~\ref{fig:rmag_completeness} below.  
Due to this incompleteness, which results in this region being quite
different from the others with respect to the abundance of $r>21$
galaxies, we have opted to drop the $\sim$1k objects 
for the dimmer (source) catalogue.  
The matches to the brighter lens catalogue were used for our
calibration; these include 
$856$ objects (80 per cent completeness).
The combination of the three PRIMUS fields adds over 3000 lens and 
5000 source calibration galaxies to our calibration sample.

Table~\ref{tab:calibcounts} shows a summary of the number of galaxies passing the cuts from each of the above calibration fields, and the remaining fraction 
after cuts described below (Sec.~\ref{ssec:obs_correction}, Sec.~\ref{ssec:photozbytemplates}) are applied. 


\section{Calibration set adjustments}
\label{sec:test_calib}

The spectroscopic calibration sample is an extremely small subset of the SDSS photometric sample
(Fig.~\ref{fig:calib_subsets}).
To ensure that conclusions derived from  these calibration samples are
applicable to the full SDSS sample, the following differences between calibration
subsamples must be addressed:
\begin{itemize}
\item The differing spectroscopic survey completeness in each
  subsample.
\item The differing SDSS observational conditions in the imaging data
overlapping each spectroscopic calibration subsample (SDSS completeness).
\item The specific large-scale structure (LSS) in each of the calibration fields
(sample variance).
\end{itemize}
We generate two calibration sets, one each for the lens and source catalogue.
The first two issues are most relevant for the source catalogue, as its faint galaxies 
are less robust to these issues.  

\subsection{Spectroscopic completeness}
\label{ssec:spectro_completeness}

\begin{figure}
\includegraphics[trim=13mm 0mm 10mm 12mm, clip, width=88mm]{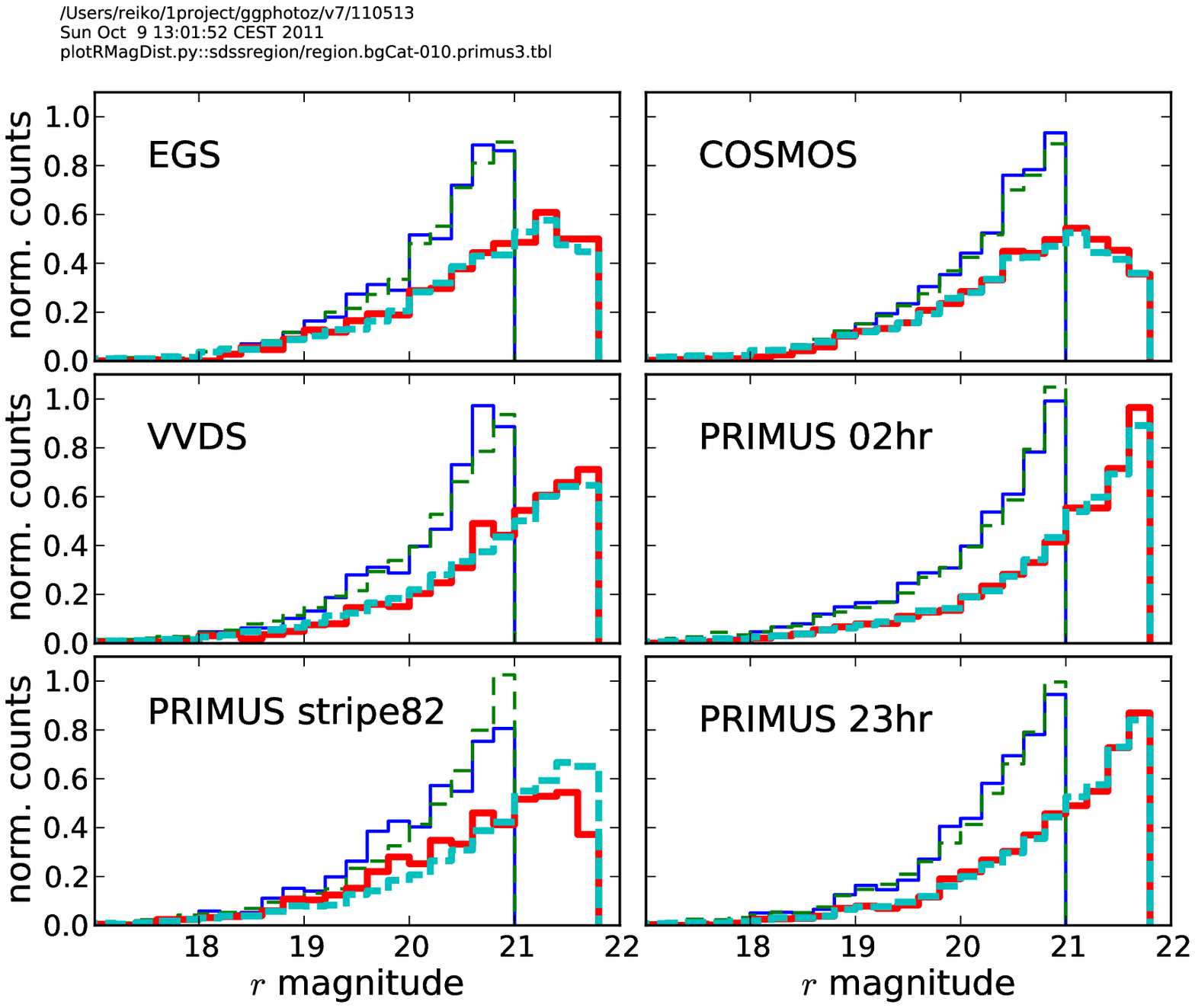}
\caption{
The $r$-band magnitude distribution of the spectroscopic calibration samples,
matched to the source ($r<21.8$, thick solid line) and photometric ($r<21$, thin solid line) 
SDSS catalogues, respectively.  
The corresponding dashed lines indicate the underlying SDSS distribution in the same area.
There is a noticeable discrepancy within $\sim0.5$ mag of the limiting
magnitude for the PRIMUS-stripe82 
spectroscopic set relative to SDSS, indicating spectroscopic incompleteness 
for the source sample at $r>21$ (see text). 
}
\label{fig:rmag_completeness}
\end{figure}

\begin{figure}
\includegraphics[trim=13mm 0mm 10mm 12mm, clip, width=88mm]{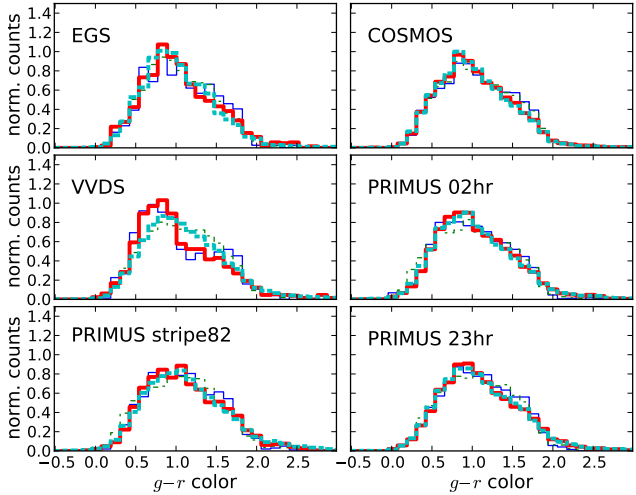}
\caption{
The $g-r$ colour distribution of the spectroscopic calibration samples,
matched to the source ($r<21.8$, thick solid line) and photometric ($r<21$, thin solid line) 
SDSS catalogues, respectively.  
The corresponding dotted lines are the underlying SDSS distribution in the same area.
Difference in the dotted and solid lines indicates colour dependence of the success rate
of the spectroscopic redshifts.  
VVDS shows some colour dependence ($g-r$) of the redshift success rate (see text).
}
\label{fig:gmr_completeness}
\end{figure}

The SDSS-matched spectroscopic calibration samples described in Sec.~\ref{ssec:spectroset}
have varying degrees of magnitude and colour completeness.
Here we address the targeting incompleteness 
of the spectroscopic set with respect to the SDSS sample. 
The completeness of the SDSS imaging data itself is
 addressed in Sec.~\ref{ssec:sdss_completeness}.



Figure~\ref{fig:rmag_completeness} shows the $r$-band magnitude distribution 
of the spectroscopic sample, matched to the source ($r<21.8$) and photometric 
($r<21$) SDSS catalogues (the thick and thin solid lines, respectively).  
The corresponding dashed lines show the underlying SDSS distribution in the same area.
The PRIMUS-stripe82 field shows a noticeable discrepancy in the number counts at 
$r>21$ between the spectroscopic redshift sample and the underlying 
magnitude distribution.  The magnitudes used for target selection 
in this field came from an earlier photometric reduction of the SDSS single-epoch 
imaging.  The earlier reduction resulted in magnitude incompleteness at the faint end,
caused by the scatter in the derived magnitudes between different reductions.
Since the PRIMUS-stripe82 source-matched catalogue also shows a lack of high redshift ($z>0.7$) 
objects which are important for source catalogue calibration, we remove this field 
from the source calibration set (but not from the lens calibration
set, since there are fewer discrepancies there).

Similarly, the thick and thin solid lines in Figure~\ref{fig:gmr_completeness} show 
the $g-r$ colour distribution of the spectroscopic sample.
Differences in the dashed (underlying SDSS distribution) and solid (distribution of the
high-quality spectroscopic redshifts) lines indicate colour dependence
of the spectroscopic redshift success rate.  
Most samples show no noticeable discrepancy in the $g-r$ colour
distributions.  We have examined all 4 colours ($u-g$, $g-r$, $r-i$, $i-z$), 
and found VVDS shows some discrepancy in the $g-r$ colour (shown), and to a
lesser degree, in $r-i$.  
Although this is alarming, we have otherwise not found anomalous 
behaviour in the VVDS subsample; the redshift distribution, the \photoz\ spectral
template type distribution, and the various derived biases are
consistent with those for the other subsamples.  Hence we keep the VVDS calibration field.
We note that the PRIMUS-stripe82 source-matched sample shows
discrepancy in the $r-i$ and $i-z$ colours, but in such a way that 
mimics the colour distribution of the brighter sample; this indicates
consistency with the incompleteness in the $r$ magnitude distribution.

\subsection{SDSS completeness}
\label{ssec:sdss_completeness}

Here we address completeness of the SDSS itself.
Different SDSS fields exhibit different $r$ magnitude distributions,
due to the varying observational conditions.  This can be seen
in the different panels of Fig.~\ref{fig:rmag_completeness}, which
have different dashed histograms (which are for the source samples in
these regions {\em before} requiring a match in the spectroscopic
data), though in the absence of other information we cannot rule out
that these variations are due to sampling variance rather than
observing conditions.  Here we will demonstrate how the non-uniform 
seeing and sky noise over the photometric survey area, 
which affects the depth to which an object can be detected, can give
rise to the differences in this figure.  
For the calibration set to fairly represent the SDSS as a whole,
we need to understand how the observing conditions in each field differ from the
median of the whole survey, and correct for any severe deviations from
typical conditions.

\subsubsection{SDSS observing conditions}
\label{ssec:sdss_obs_conditions}

\begin{figure}
\includegraphics[width=84mm]{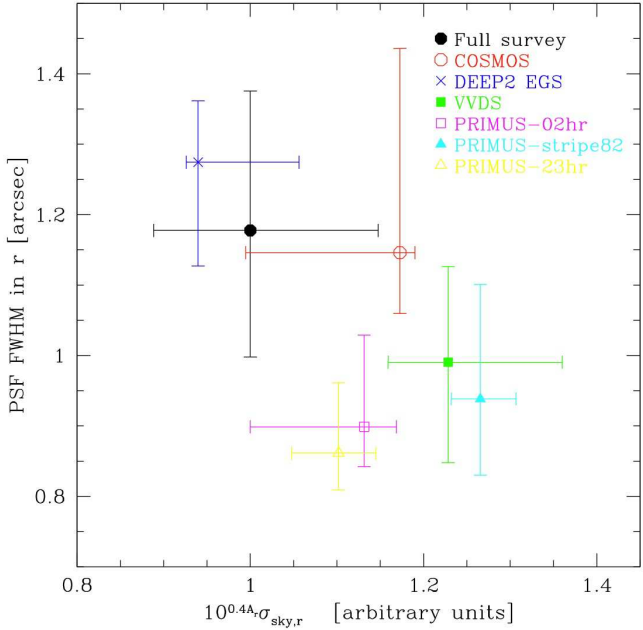}
\caption{
SDSS $r$-band observation conditions (sky noise, as described in the text, and seeing)
for each of the calibration fields and the SDSS full survey. 
}
\label{fig:sdss_observation_condition}
\end{figure}

Figure~\ref{fig:sdss_observation_condition} shows the SDSS observation
conditions in $r$-band, for the source galaxies in each of the
calibration fields and in the full SDSS survey.  Here we have used
galaxies that are well-resolved to trace the observing conditions,
which means that we will be skewed towards better observing conditions
(as compared to if we had done the calculation using pure areas).
This methodology explains why the seeing is noticeably better than the
typical value that is commonly used for SDSS, 1.4$^{\prime\prime}$.  
Points indicate a median value, and error bars indicate the 68th percentile.
To calculate sky noise (horizontal axis), we approximate the Poisson
noise due to the 
sky and dark current as 
\begin{equation}
\sigma_\text{sky}(\text{counts}) = \sqrt{\frac{\text{sky}}{\text{gain}} + \text{dark current}}.
\end{equation}
This is a reasonable approximation for galaxies with 
$r\gtrsim 20$, where the Poisson noise due to the galaxy flux
itself is negligible.  We then convert this to the sky noise in
nanomaggies using the calibration from ubercal
\citep{padmanabhan_etal:2008}; this noise in nanomaggies 
determines the  $S/N$ for a galaxy with a given size and flux, at fixed seeing
and galactic extinction $A_r$.  Then, we multiply the sky noise
estimates by $10^{0.4A_r}$; in reality, of course, extinction
modulates the flux, but this has the same effect on the S/N as
increasing the sky noise by this factor.   Finally, we have
divided out by the median value of $10^{0.4A_r}\sigma_\text{sky}$ for the survey.  
When the sky noise is large, more galaxies at a given flux will fail the $S/N>5$
object detection filter; this is also true for worse seeing, since that filter is imposed within a PSF
rather than using the entire galaxy flux.  

The $r$-band PSF FWHM in arcsec is shown on the vertical axis.  
In poor seeing, the observed magnitudes
are noisier (because the galaxy is spread out over more pixels and
has greater sky noise contribution), and
star-galaxy separation is more challenging (more galaxies get
classified as stars). 

As shown, the observing conditions in the calibration fields are not
typical of the full SDSS DR8 sample.  
The EGS is within the $1\sigma$ range of
observing conditions for both seeing and sky noise.  The COSMOS data
have relatively high sky noise for more than half the area, which will
lead to reductions in galaxy $S/N$
of $\sim 20$ per cent compared to that for similar galaxies in the EGS.
PRIMUS-02hr and PRIMUS-23hr are marginally within the typical region for sky
noise, and VVDS and PRIMUS-stripe82 are at higher sky noise; but all four fields
have significantly better seeing than that of the full SDSS.  The reason for this
trend is that all of these samples are on stripe 82, which has many
observations.  Since we select towards observations with better seeing
(preferable for lensing) when creating the source catalogue, we get atypically good seeing.  The fact that the extinction is on
the higher end for these regions increases the effective sky noise,
though it is not the only (or even the main) reason why these
calibration samples have higher sky noise than is typical.
As we will see in the next section,
to lowest order we expect that the changes in sample properties due
to having atypically good seeing will oppose (and even outweigh) the changes in sample
properties due to atypically high sky noise for these stripe 82 samples.

\subsubsection{Effect on number counts}
\label{ssec:obs_cond_effect}

\begin{figure}
\includegraphics[width=84mm]{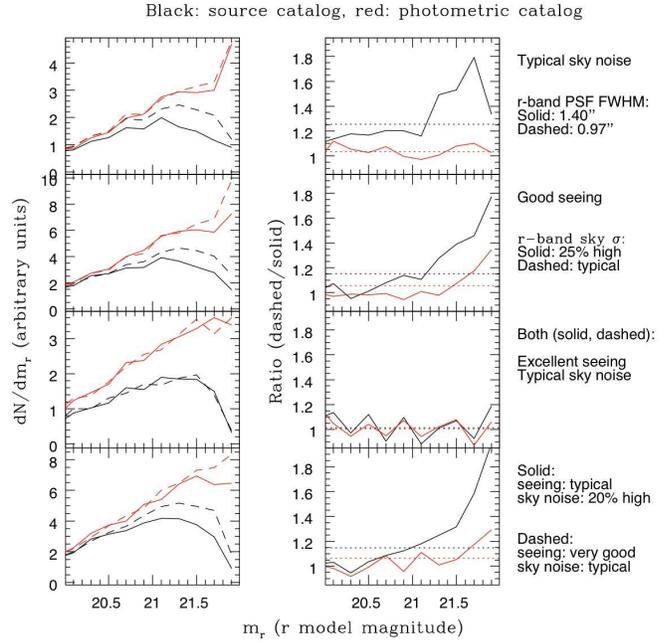}
\caption{
An illustration of how observing conditions can modulate the observed
galaxy number densities, based on comparisons of pairs of runs with
identical area coverage but different observing conditions in stripe
82.  Each row utilizes a different pair of runs with conditions
given on the right.  Black lines are for the source catalogue, red
lines are for the full photometric catalogue (no lensing
selection).  {\em Left column:} Histogram of $r$-band model
magnitudes, with the different colours indicating which catalogue was
used, and different line types indicating the observing conditions as
labeled on the right-hand side.  {\em Right column:} Solid lines show
the ratio of the histograms with different observing conditions from
the left column, always putting the dashed histogram in the numerator.
The dotted lines show the ratio of the total number of galaxies in the
region as defined using the full sample (averaged over magnitudes).
}
\label{fig:observing_condition_effect}
\end{figure}

In Fig.~\ref{fig:observing_condition_effect}, we illustrate the effect
of observing conditions on the overall counts and  magnitude distribution of galaxies, which is
helpful in understanding the behaviour of the galaxy counts in the
different calibration samples.  We make this comparison by selecting
pairs of runs overlapping the PRIMUS-02hr and PRIMUS-23hr regions, with
different observing conditions but exactly the same area coverage.
Thus, 
any differences in the galaxy apparent magnitude distributions in the
pairs of runs are due to the observing conditions,
rather than intrinsic differences in the galaxy populations.  The
four rows correspond to four different scenarios, with r model magnitudes on the horizontal axis 
and number counts on the vertical axis.  At left are the
counts (for both fields, source and photometric lens catalogues).
At right the ratio between counts for the runs with two different observing conditions is shown.   In the top row,
results are given for nearly identical (within 10 per cent) sky noise, but 45 per cent difference in the seeing.
(This is done by comparing runs that are $\sim 20$ per cent better and worse than the survey
median seeing.)  The second row corresponds to seeing that is nearly identical
(within 10 per cent) and a bit better than the survey median, but sky noise
that differs by 30 per cent.  The third row shows nearly identical conditions
for the seeing and sky noise, to illustrate the magnitude of the
differences that can occur just due to Poisson noise.  The fourth row shows
the case of two effects going in the same direction: 25 per cent difference in 
seeing and 15 per cent difference in sky noise, where one run is better in both 
cases (better seeing and lower sky noise). 

For the photometric catalogue, the effect of seeing on the galaxy number
counts is small ($<5$ per cent, as shown in the top row) and nearly
constant over all magnitudes.  This change in counts may be due predominantly to
the star-galaxy selection.  For the source catalogue, which has apparent
size selection, the effect of seeing is larger (as we would expect), and varies
more strongly with magnitude, from a nearly constant 20 per cent below $r<21$
to typically 50 per cent above $r>21$.  

The signature of sky noise, shown in the second row, is somewhat different.
It preferentially reduces the number counts at fainter magnitudes
for the purely flux-limited galaxy sample. The overall effect of 5 per cent 
is actually dominated by the fainter parts of the catalogue, with almost no effect below $r<21.4$ and a gradually increasing effect at fainter magnitudes, up to 35 per cent.  In contrast, for the source catalogue, the average difference is more significant (15 per cent) and becomes noticeable at brighter magnitudes, from $r>\sim 20.8$.
The third row shows that for identical observing conditions, the samples that are selected in both 
the photometric and source catalogue are statistically identical, with $<5$ per cent fluctuations in the magnitude histograms.  Examination of the actual galaxy samples indicates significant noise, with fainter galaxies being scattered in and out of the sample, such that the fraction of galaxies in both catalogues is $\sim 70$ per cent.  Finally, the fourth row is not at all surprising considering the first and second: we see both the effects of seeing and of sky noise.

From this analysis we see that
the flux-limited photometric lens sample, with $r<21$, is not strongly
affected by the changes in observing conditions, because 
these effects are only very significant ($>5$ per cent) in objects fainter than $r\sim 21.4$.  
In contrast, the source number density is strongly affected by observing conditions.
We now turn to
the corrections we make to account for these effects in our
calibration samples, to make them more representative of the full SDSS
area.

\subsubsection{Corrections for observation conditions}
\label{ssec:obs_correction}

\begin{figure}
\includegraphics[trim=12mm 0mm 10mm 15mm, clip, width=88mm]{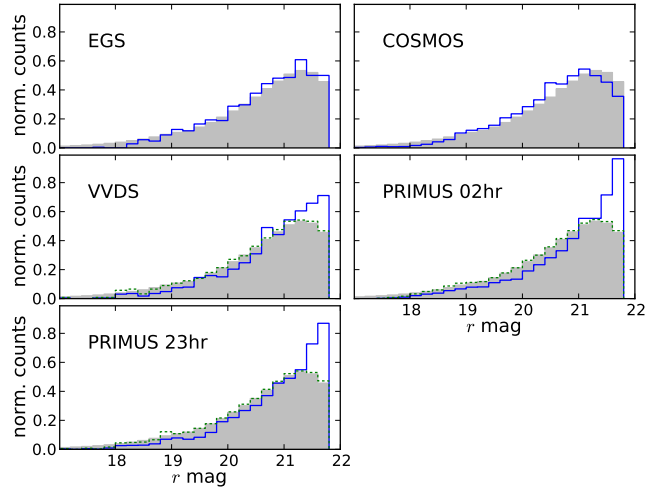}
\caption{
The $r$-band magnitude distribution of the source catalogues for each spectroscopic subsample
(line histograms), in contrast to the SDSS distribution (grey block histograms).
The differing SDSS observing conditions cause variable magnitude completeness 
near the limiting magnitude $r\sim21.8$.
We adjust the galaxy population in VVDS, PRIMUS-02hr, and PRIMUS-23hr fields to
match the SDSS completeness by systematically trimming the lowest resolution objects
(dotted lines), such that the resulting $r$ magnitude distribution matches that of SDSS
(see text).
}
\label{fig:compare_rmag_completeness}
\end{figure}

Figure~\ref{fig:compare_rmag_completeness} explicitly compares the magnitude distribution
for each of the source calibration samples to that of the whole of SDSS.
Objects in the fields that overlap SDSS stripe 82 (VVDS, PRIMUS-02hr and PRIMUS-23hr)
tend to have high completeness near the limiting magnitude; this is because
there are many repeat observations in this stripe, and the source
catalogue reductions choose the observation of each galaxy that has
the best seeing.

Since stripe 82 only covers a small fraction of the SDSS footprint
($\sim$3 per cent) and is not representative of the whole of SDSS, 
we would like to correct for the excess faint objects.
These objects are removed as follows: (1) we bin the galaxies by their $r$ magnitudes, 
and (2) for each magnitude bin, we remove the lowest apparent
resolution galaxies until the 
resulting magnitude histogram matches that from all of SDSS.  This process removes
$\sim 40$ per cent of the objects in a given field.
(Figs.~\ref{fig:rmag_completeness}  and \ref{fig:gmr_completeness} do
not include this cut.)

The COSMOS field has poor sky noise, and in general has unusually bad observational conditions
compared to the rest of SDSS (e.g., a third of the COSMOS field is not
classified as 
photometric in SDSS, so we have eliminated that portion of the field).   
This leads to a deficit of faint galaxy detections; however, we cannot artificially add objects 
that are missing from the catalogues.  
Hence, when using the COSMOS source calibration set, we have applied empirical corrections 
to the derived values (see Sec.~\ref{ssec:source_lensbias}).

\subsection{Sample variance}
\label{ssec:sample_variance}

\begin{figure}
\includegraphics[trim=12mm 0mm 10mm 15mm, clip, width=88mm]{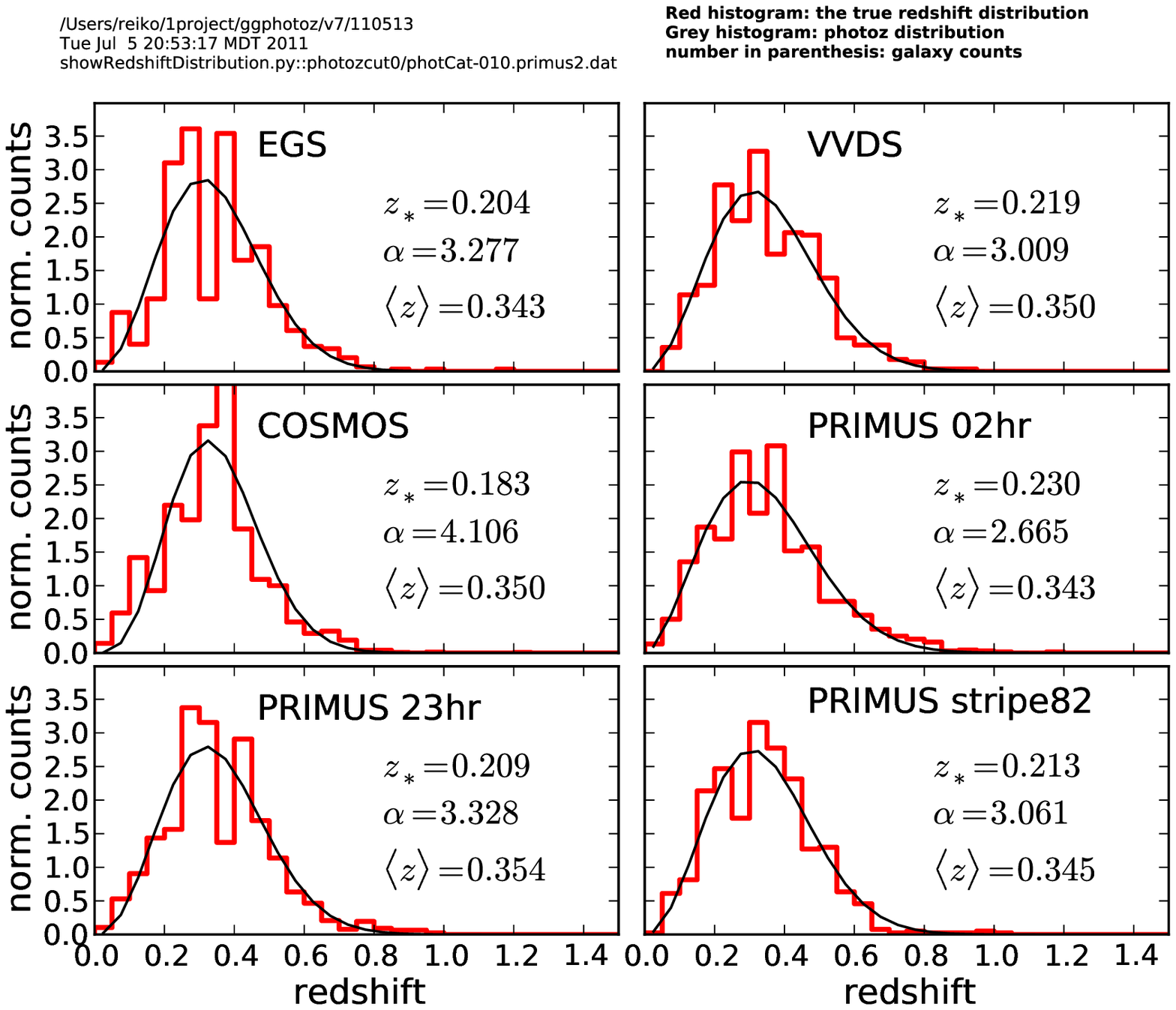}
\caption{
The observed redshift distributions of the lens calibration subsamples
are shown as thick line histograms.
The fits to the underlying redshift distribution (thin lines), along
with the best-fitting parameters, are also shown.
}
\label{fig:lens_specz_distribution} 
\end{figure}

\begin{figure}
\includegraphics[trim=12mm 0mm 10mm 15mm, clip, width=88mm]{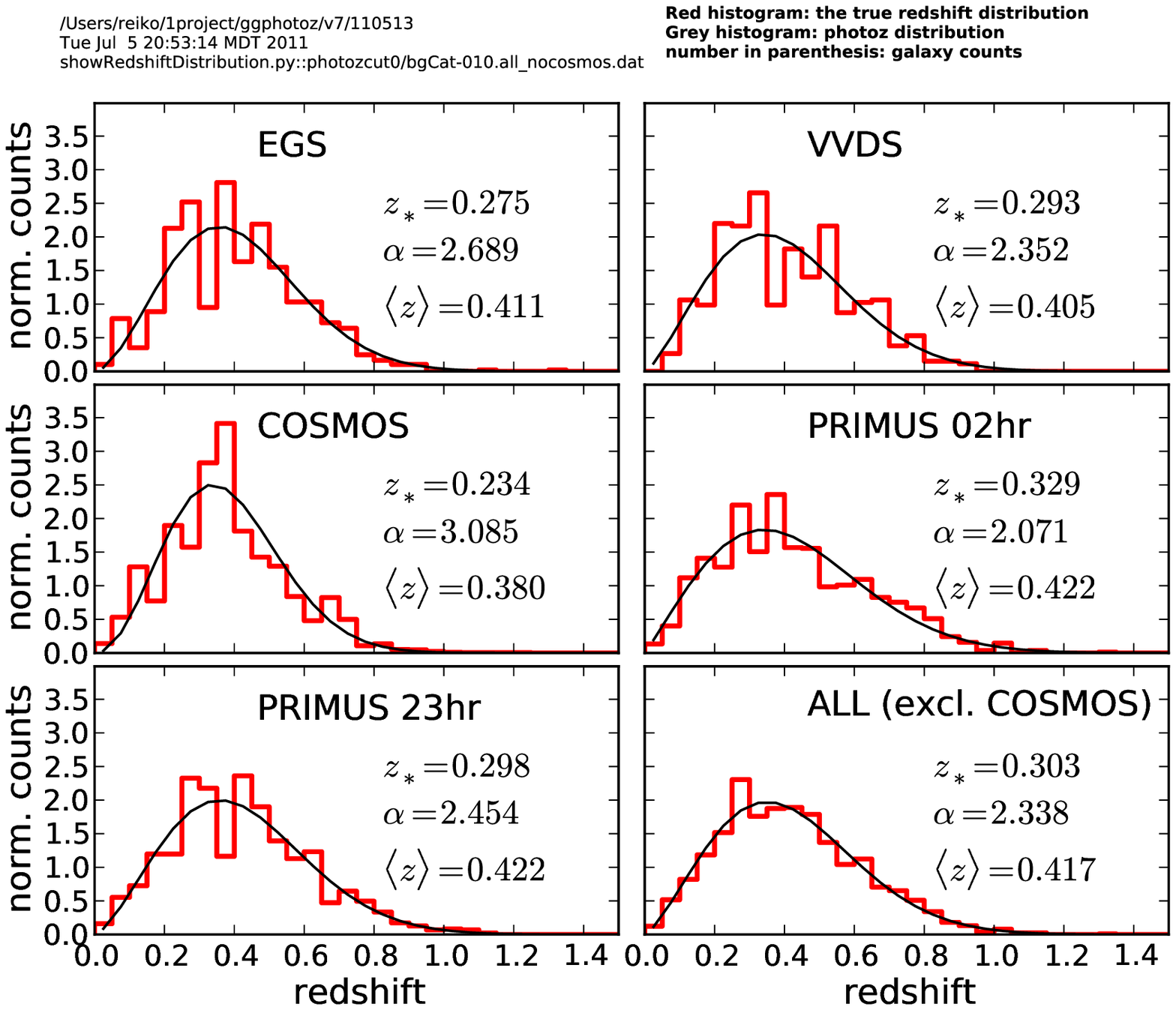}
\caption{
Analogous to Fig.~\ref{fig:lens_specz_distribution}, except these are for 
the source calibration subsamples, along with the combination of all the source calibration
subsamples excluding the COSMOS field (see text).
}
\label{fig:source_specz_distribution} 
\end{figure}

\begin{table}
\caption{Fitting parameters for redshift distributions of the subsamples.
The source samples for VVDS, PRIMUS 02hr and PRIMUS 23hr have the seeing cuts
(Sec.~\ref{ssec:obs_correction}) applied; the redshift distribution parameters before 
and after \photoz\ cuts (Sec.~\ref{ssec:photozbytemplates}) are listed.
}
\label{tab:zfitparams}
\begin{tabular}{lr rr rr}
\hline
Sample & $N_{\rm gal}$ & $z^*$ & $\alpha$  & $\langle z\rangle$ \\
\hline
\multicolumn{5}{l}{Photometric (lens) sample with r$<$21} \\
EGS  &  639 & 0.209 & 3.050 & 0.336 \\
VVDS &  643 & 0.224 & 2.710 & 0.338 \\
COSMOS & 3281 & 0.207 & 3.139 & 0.339 \\
PRIMUS 02hr & 1508 & 0.243 & 2.325 & 0.334 \\
PRIMUS 23hr & 1677 & 0.220 & 2.987 & 0.349 \\
PRIMUS stripe82 & 856 & 0.223 & 2.718 & 0.336 \\
\hline
\multicolumn{6}{l}{Photometric (lens) sample with r$<$21 and \photoz\ quality cuts}\\
EGS  & 593 & 0.204 & 3.277 & 0.343 \\
VVDS & 564 & 0.219 & 3.009 & 0.350 \\
COSMOS & 2959 & 0.183 & 4.106 & 0.350 \\
PRIMUS 02hr & 1357 & 0.230 & 2.665 & 0.343 \\
PRIMUS 23hr & 1547 & 0.209 & 3.328 & 0.354 \\
PRIMUS stripe82 & 786 & 0.213 & 3.061 & 0.345 \\
\hline
\multicolumn{5}{l}{Source sample with r$<$21.8 (includes seeing cuts)}\\
EGS  & 1060 & 0.289 & 2.399 & 0.404 \\
VVDS &  609 & 0.253 & 2.101 & 0.391 \\
COSMOS & 4764 & 0.253 & 2.560 & 0.368 \\
PRIMUS 02hr & 1834 & 0.344 & 1.854 & 0.411 \\
PRIMUS 23hr & 1896 & 0.308 & 2.257 & 0.416 \\
\hline
\multicolumn{6}{l}{Source sample with r$<$21.8 and \photoz\ quality cuts}\\
EGS  & 968 & 0.275 & 2.689 & 0.411 \\
VVDS & 527 & 0.293 & 2.352 & 0.405 \\
COSMOS & 4293 & 0.234 & 3.085 & 0.380 \\
PRIMUS 02hr & 1646 & 0.329 & 2.071 & 0.422 \\
PRIMUS 23hr & 1737 & 0.298 & 2.454 & 0.422 \\
\hline
\end{tabular}
\end{table}

Each individual field has distinct large-scale structure due to galaxy
clustering.
Thus, the observed $\rmd N/\rmd z$ in each field differs from the true
underlying redshift distribution $\rmd
N/\rmd z$,
which is unknown but is expected to be smooth.
We model this smooth distribution by fitting to a functional form 
(our bias analysis does not depend significantly on the detailed form of the curve, as discussed in
\citealt{mandelbaum_etal:2008})
\begin{equation}
\frac{\rmd N}{\rmd z} \propto \left(\frac{z}{z_*}\right)^{\alpha - 1} 
   \exp\left[-\frac{1}{2}\left(\frac{z}{z_*}\right)^2\right]
\label{eqn:fitting_formula}
\end{equation}
which has a mean redshift of 
\begin{equation}
\langle z \rangle = \frac{\sqrt{2} \;z_*\; \Gamma[(\alpha+1)/2]}{\Gamma[\alpha/2]}.
\end{equation}
Figures~\ref{fig:lens_specz_distribution} and \ref{fig:source_specz_distribution}
show the true redshift distributions for each of the lens and source 
calibration samples, along with their fitted curves and parameters.
Following Sec.~4.2 of \citet{mandelbaum_etal:2008}, 
the histograms are $\chi^2$-fit to Eq.~\eqref{eqn:fitting_formula} with  
flat weights while constraining
the area under the curve to equal to the galaxy count $N_{\rm gal}$;
the redshift histograms of binning width $\Delta z=0.05$ were bootstrap resampled to 
obtain LSS error estimates to the fitting curve.
The distributions shown here were derived after imposing cuts to
correct for different seeing in the calibration samples on stripe 82,
as discussed in Sec.~\ref{ssec:obs_correction}.  When further \photoz\ quality cuts
are applied (Sec.~\ref{ssec:photozbytemplates}), the distribution shifts somewhat; 
the fit parameter values for both cases are listed for each catalog and subsample
in Table~\ref{tab:zfitparams}.

The errors in the fits to the redshift distribution for each sample come primarily from
sampling variance; thus, when fitting to the overall calibration
sample redshift distribution, these fluctuations are reduced (more
significantly than one would expect fluctuations due to Poisson noise
to be reduced). 
The significant smoothing out of LSS can be seen in the bottom right panel of 
Fig.~\ref{fig:source_specz_distribution}, where we plot the sum of all 
source subsamples, excluding the COSMOS region
(this region was excluded due to its difference in sky noise which
should result in a different intrinsic redshift distribution). 
This curve is our best estimate for the source catalogue redshift distribution;
a similar smoothing in LSS is seen for the lens sample, for the sum of all 6 subsamples.
These figures show that the upper limit in redshifts is around $z\sim0.8$ ($z\sim1.0$) for the 
lens (source) catalogues, with a mean redshift of $0.34$ ($0.41$).
We note that the parameters $z_*$ and $\alpha$ are degenerate to some extent
(small $\alpha$ brings the peak to higher redshifts, while high $z_*$ spreads out the 
overall distribution), and hence there is considerably more uncertainty in the parameters
in $[z_*, \alpha]$ than in the mean redshift $\langle z \rangle$.
The fit parameter errors can be estimated by modulating the redshift bin counts
to simulate LSS \citep{mandelbaum_etal:2008}, or by looking at the variation in the fit 
values of the different subsamples.  

\subsection{Other incompleteness}
\label{ssec:deficient_redshift_bins}

The EGS matched samples have a deficiency of $z\sim0.3$ galaxies, which is due to
instrumental limitations of the survey\footnote{For blue galaxies, $[OI\!I\!I]$ 
and $H\alpha$ lines are both off the spectrum 
(particularly in the vignetted corners where the spectral coverage 
is reduced), and for red galaxies, there is difficulty in confirming 
a redshift when there are no strong spectral features available, but only the 
forest of absorption lines.  Both of these problems turn out to occur at similar 
redshifts in the DEEP2 EGS survey.} (Newman et~al\@. 2011 in~prep., priv. comm.).
Such a deficiency cannot easily be remedied; however,
we find that, at this redshift, this deficiency does not affect the
derived lensing signal calibrations when using source \photoz, which is more sensitive to the distribution of
high-redshift ($z>0.6$) objects.  However, this does affect the
calibration for lens \photoz,
and care must be taken when the EGS lens calibration subset is used on its own
(Sec.~\ref{ssec:lens_lensbias}).

\section{Photo-$z$ Method}
\label{sec:photoz_method}

We chose the template-based ZEBRA code \citep{feldmann_etal:2006} to
estimate our \photoz's based upon the SDSS $ugriz$ bands.
There are several public SDSS \photoz\ catalogues for the full photometric sample, 
all of which are based on training-set methods: 
two by \citet{oyaizu_etal:2008}, available in the DR8 {\tt CAS} 
database\footnote{\tt http://skyserver.sdss3.org/dr8/en/},
one by \citet{budavari_etal:2000} in SDSS DR7 \citep{sdssdr7:2009} 
{\tt CAS} database\footnote{\tt http://casjobs.sdss.org/dr7/en/},
and another by 
\citet{cunha_etal:2009}\footnote{\tt http://www.sdss.org/dr7/products/value\_added/index.html}.

These \photoz\ methods were included among those tested in \citet{mandelbaum_etal:2008}, for 
source galaxy redshifts.  However, as those estimates were based on earlier
Data Releases (5 and 6), including more preliminary \photoz\
estimation procedures and photometry that lacked ubercalibration, the
results of the tests in \citet{mandelbaum_etal:2008} do not necessary
apply to these new DR7 and DR8 versions. 
In Appendix~\ref{app:other_photozs}, we 
compare the \photoz's in those publicly available SDSS DR8 catalogue,
and give our reasons for choosing to use ZEBRA, another template-based
method that will be described below.

Template-fitting \photoz\ algorithms operate, broadly speaking, as follows:
Given a set of spectral energy distribution (SED) templates, the SEDs
are projected onto passbands shifted on a grid of redshifts, from which
the predicted photometry, and hence colour, for each case (template type and redshift) 
is calculated.  The observed colours are then compared against the model
magnitudes, with a $\chi^2$-minimisation over all bands used to determine the best-fit redshift 
and spectral type.  
 The choice of templates is crucial for obtaining quality \photoz, so
 we will begin by describing our choice of templates.

\subsection{Templates}

\begin{figure}
\includegraphics[trim=2mm 0mm 10mm 10mm, clip, width=88mm]{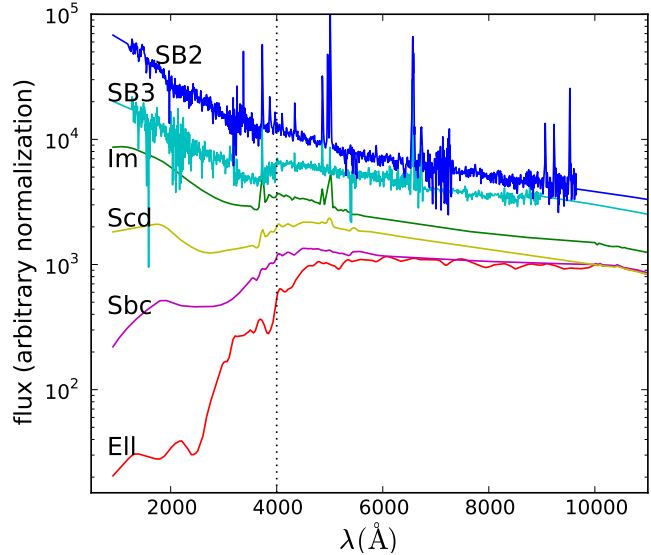}
\caption{
The six SED templates used in the ZEBRA photometric redshifts.
Four templates (labeled Ell, Sbc, Scd and Im) are observed spectra from
\citet{coleman_etal:1980}, and two (SB2 and SB3) are synthetic starburst 
spectra \citep{kinney_etal:1996}.
Five templates were interpolated between the existing 6 templates, 
for a total of 31 templates which were used for ZEBRA \photoz\ estimation.
}
\label{fig:zebra_templates}
\end{figure}

Templates can be obtained from observation, or generated from stellar population
synthesis (SPS) models.  
We choose the set of six SED templates as described in \citet{benitez:2000}, and used in
ZEBRA \citep{feldmann_etal:2006}.  They are shown in Figure~\ref{fig:zebra_templates}. 
The set consists of four
templates from \citet[CWW hereafter]{coleman_etal:1980}, which are SEDs observed
across a wavelength baseline of 1400\AA\ $<\lambda<10000$\AA\ in the local ($z\sim0$) universe;
they are supplemented by two synthetic blue spectra, using the GISSEL synthetic models
\citep{bruzual_charlot:1993}.
While each of these templates have been linearly extrapolated into the UV and NIR wavelengths
based on different synthetic models\footnote{The SED extensions used here differ from
those of \citet{ilbert_etal:2006}, who use the same template sets, but whose extensions are updated
with the more recent \citet{bruzual_charlot:2003}.}, these extensions are largely irrelevant
to the optical filter set that we use (3000\AA\ $< \lambda < 10000$\AA) for the redshift
range relevant for our sample ($z<1.5$).  The CWW templates have been widely employed
and are known to generate reliable photometric redshift estimates, although training and
tweaking the templates improves redshift estimates \citep{ilbert_etal:2006, feldmann_etal:2006}.

In particular, \citet{ilbert_etal:2006} find that these templates work
reasonably well for VVDS data,
with a \photoz\ scatter of $\sigma_{\Delta z/(1+z)}\sim 0.029$ given deep
photometry covering a similar range of wavelength to ours
($u^*g'r'i'z'$ and $BVRI$, with additional $JK$ bands for 13 per cent
of galaxies).   
With SDSS photometry, we expect a larger 
scatter because of the shallower photometry and lack of extra bands.\footnote{
We tested other template sets, such as the Poggiani templates used by \citet{abdalla_etal:prep} and
Polletta templates generated by GRASIL \citep{ilbert_etal:2009}.
We find the CWW + Kinney templates to be sufficient for the SDSS photometric samples.
\citet{abdalla_etal:prep} find that the Poggiani template set works
well with the LRG samples.  
We confirmed their results for the redshift range $0.4<z<0.8$ and the galaxy types
in their sample; however, it was not as successful as our fiducial choice 
outside of that redshift range and colour selection.
The Poletta templates 
were required for good \photoz\ estimates in \citet{ilbert_etal:2009} 
due to their use of IR photometry; in our case, with only optical data, 
we find no significant difference from the default CWW+Kinney templates.}

\subsection{ZEBRA options}
For the different template-fitting \photoz\ codes, their \photoz\ accuracy are very similar
if the same set of templates is used, since most methods use 
essentially similar basic procedures of $\chi^2$ minimization
\citep{PHAT:2010}.  
However, in addition to template choice,
ZEBRA offers several options that enhances the \photoz\ calculations.
Priors on the redshift distribution or other conditions may be applied when
calculating $\chi^2$.  Additional features, such as iterative photometry 
self-calibration, template optimisation, $k$-correction tables, and
different modes of \photoz\ and template type determination, are also available.
We detail our choices of such ZEBRA options here.

\subsubsection{ZEBRA parameters}

We ran ZEBRA in the maximum-likelihood (ML) mode.  This was chosen over the 
Bayesian mode (BP), which allows for a Bayesian prior on the redshift distribution, 
per template type and redshift range.  Although the latter method
removes the \photoz\ bias to some extent, we find that it significantly adds to the scatter in
\photoz, which is the quantity of most concern for our application.  
Because the original CWW templates are known to generate \photoz's with bias
\citep[Fig.~5 of][]{feldmann_etal:2006}, applying a ``correct'' redshift distribution prior
(the estimated redshift distribution of the given sample) will decrease the bias
while increasing the scatter.

We use the following set of parameters to obtain our \photoz's.
The SED templates were interpolated with 5 new templates between each
of the 6 given templates, resulting in a total of 31 templates. 
The redshifts are allowed to vary in steps of 0.0025 from 0 to 1.5, 
with no smoothing of the filter bands.
We have applied a prior of \photoz\ $<1.5$, which is reasonable 
for single-epoch SDSS photometry (most galaxies within our magnitude
limit are at $z<1.0$).
Similarly, we do not correct for the Lyman-$\alpha$ IGM absorption 
\citep{madau:1995, meiksin:2006} since it is only relevant for objects at $z>3$.

\subsubsection{ZEBRA self-calibration}

In principle, the template-fitting may benefit from use of a \specz\ training set
to calibrate the zero-point magnitude of the photometry; here, this
procedure is unnecessary as we find no offset for SDSS.
This is as expected given the
excellent photometric calibration \citep{padmanabhan_etal:2008}.
The SED templates can be modified based on a \specz\ calibration set, but
we also find this template modification procedure unnecessary.
While template calibration have been shown to work with samples of better photometry
\citep{ilbert_etal:2006, feldmann_etal:2006},
our attempts at template modification suggested that we were only
fitting to the photometric noise, as an improvement to the templates
and \photoz\ error based on 
one calibration set did not translate to less scatter in the \photoz\
computed with other calibration sets.    Moreover, the modifications
to the templates did not look remotely physical.  
The difference between our finding and the improvements found in \citet{feldmann_etal:2006}
most likely originates from the low signal-to-noise of our photometry.


\subsection{$k$-correction}
\label{ssec:zebra_kcorrection}

ZEBRA, as a template fitting method, allows estimates of
absolute magnitudes and stellar masses.  To obtain these,
we have utilised the $k$-correction table (a list of theoretical magnitudes
for each SED template at all redshifts; Eq.~\eqref{eqn:kcorrect_tabular} below) 
generated by ZEBRA.  
This table is then modified according to the procedures below, to 
allow versatility in our $k$-corrections.  Factors of 
$\log_{10}(1+z)$ that appear with different signs and/or prefactors in
previous papers on the
topic of $k$-correction are elucidated here. 

$k$-correction encapsulates the magnitude corrections beyond the simple distance modulus,
such that
\begin{equation}
  M_Q = m_{R} - DM(z_{\rm gal}) - K(z_{\rm gal},\;T,\;R,\;Q).
\end{equation}
Here, $M_Q$ is the rest-frame absolute magnitude in filter $Q$,
$m_{R}$ is the observed magnitude in band $R$,
$DM(z_{\rm gal})=5\log_{10}(D_L/10{\rm pc})$ is the distance modulus (where 
$D_L=D_L(z_{\rm gal})$ is the luminosity distance in parsecs), and 
$K(z_{\rm gal},T,Q,R)$ is the $k$-correction factor.  
The $k$-correction depends on the galaxy redshift $z_{\rm gal}$, template type $T$, the
band $R$ in which the magnitude is observed, and the desired band $Q$ for
the absolute magnitude.  
To minimise the degree of $k$-correction in a sample, $Q$ is often chosen such that 
it is the $R$ band filter redshifted to the sample median redshift $z_0$:
i.e., $Q=^{z_0}\!\!\!R$, or $Q(\lambda(1+z_0)^{-1})=R(\lambda)$.  Then 
$k$-corrections are calculated as \citep{hogg_etal:kcorrect, blanton_roweis:2007}
\begin{eqnarray}
\label{eqn:kcorrect_define}
 K(z_{\rm gal},\;T,\;R,\;Q) &=& 2.5\log_{10}(1+z_{\rm gal}) \\
&& + \; \hat{m}(z_{\rm gal},\;T,\;R) - \hat{m}(z=0,\;T,\;Q)
\nonumber 
\end{eqnarray}
where ZEBRA tabulates the AB magnitudes $\hat{m}$ for an array of redshifts $z_{\rm gal}$:
\begin{equation}
\hat{m}(z_{\rm gal},\;T,\;R) = -2.5 \log_{10} \left[
   \frac{\int d\lambda \;\lambda \;F_\lambda^{T}[\lambda(1+z_{\rm gal})^{-1}] R(\lambda)}
        {\int d\lambda \;\lambda \;g_\lambda^{AB}(\lambda) R(\lambda)}
\right].
\label{eqn:kcorrect_tabular}
\end{equation}
Here the quantity in square brackets is the flux ratio that defines the AB magnitude. 
Since the observed SED output of a galaxy of type $T$ at redshift $z_{\rm gal}$ is
\begin{equation}
(1+z_{\rm gal})^{-1}F_\lambda^{T}[\lambda(1+z_{\rm gal})^{-1}]\,d\lambda
\end{equation}
where $F_\lambda^{T}(\lambda)\,d\lambda$ is the rest-frame SED, we see that 
the $2.5\log_{10}(1+z_{\rm gal})$ factor which appears in the first term of
Eq.~\eqref{eqn:kcorrect_define} is from correction to the tabulated magnitude
$\hat{m}(z_{\rm gal})$ for the loss in photon energy (flux) due to redshift.
The reference SED relative to which AB magnitudes are defined is $g_\lambda^{AB}(\lambda)$,
and $[\lambda\;g_\lambda^{AB}(\lambda)] \propto \lambda^{-1}$, since 
$g_\lambda^{AB}$ is proportional to $\lambda^{-2}$.  
The arbitrary normalisations in both $g^{AB}$ and $F_\lambda$
cancel out when we take the difference of $\hat{m}$.

If $Q$ is $R'$ redshifted by $z_0$, such that $R'$ is one of the
existing bands (i.e.,
$Q(\lambda(1+z_0)^{-1})=R'(\lambda)$), then the tabulated values can also be used to obtain
$\hat{m}(z=0, T, Q=^{z_0}~\!\!\!\!\!R')$:
\begin{eqnarray}
\hat{m}(0,\;T,\,^{z_0}\!R') &\equiv& -2.5 \log_{10} \left[
   \frac{\int d\lambda \;\lambda \;F_\lambda^{T}(\lambda) Q(\lambda)}
        {\int d\lambda \;\lambda^{-1} \;Q(\lambda)} \right] \nonumber \\
&=& -2.5 \log_{10} \left[
   \frac{\int d\lambda' \;\lambda' \;F_\lambda^{T}[\lambda'(1+z_0)^{-1}] R'(\lambda')}
        {(1+z_0)^2\int d\lambda' \;\lambda'^{-1} R'(\lambda')} \right] \nonumber \\
&=& \hat{m}(z_0,\;T,\;R') + 5\log_{10}(1+z_0)
\end{eqnarray}
where $\lambda'\equiv\lambda (1+z_0)$.
The $k$-correction from the ZEBRA tabulated values are then
\begin{eqnarray}
K(z_{\rm gal},\;T,\;R,\;^{z_0}\!R') &=& 
2.5\log_{10}(1+z_{\rm gal}) - 5\log_{10}(1+z_0) \nonumber \\
&& + \;\hat{m}(z_{\rm gal},\;T,\;R) - \hat{m}(z_0,\;T,\;R') 
\label{eqn:kcorrect}
\end{eqnarray}
where $R$ and $R'$ can be any of the available filter bands.  
For the special case where $R=R'$ and $z_{\rm gal}=z_0$, we have
\begin{equation}
K(z_0,\;T,\;R,\;^{z_0}\!R) = - 2.5\log_{10}(1+z_0)
\end{equation}
which does not depend on the template type $T$ by construction, but is non-zero.
Since the SED type $T$ is uncertain in all \photoz's, the observed band $R$ 
should be chosen such that it maximally overlaps with $^{z_0}\!R'$ 
to minimize the $T$-dependence of the $k$-correction.





\section{Photo-$z$ accuracy}
\label{sec:photoz_results}

In this section, we report the ZEBRA \photoz\ accuracy on the SDSS DR8
photometry, using all of the spectroscopic calibration subsamples together.  
Since most of our galaxies are near the magnitude limit where the photometry is noisy,
the derived \photoz's have a relatively large uncertainty.
We consider dependence of \photoz\ scatter on template, and use this information 
to apply \photoz\ quality cuts.
We have not found a significant correlation of the $\chi^2$ values to the actual \photoz\ errors.


\subsection{Photo-$z$ errors by template types}
\label{ssec:photozbytemplates}

\begin{figure}
\includegraphics[trim=2mm 10mm 8mm 20mm, clip, width=88mm]{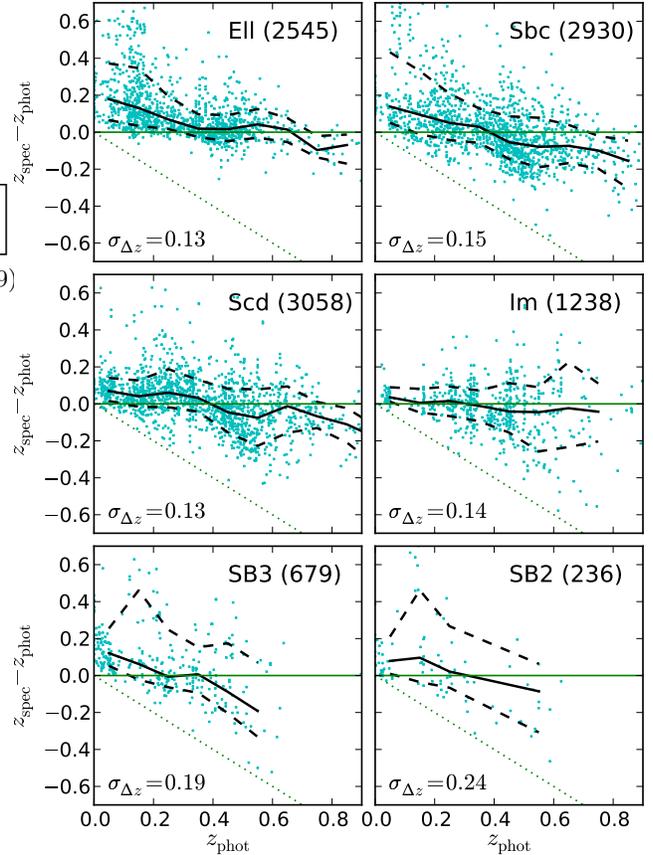}  
\caption{
Photometric redshift errors in the source catalogue by template types
(Ell, Sbc, Scd, Im, SB3, and SB2), as a function of the photometric redshift (the observable).
The number in parenthesis indicates object counts.
The median (bold solid line) and the 16 and 84 percentile range (bold dashed line)
of $\Delta z$ is also shown.
The \photoz\ scatter is given as the standard deviation of $\Delta z$ 
{\em from the median}.  Only 50 per cent of the points are shown for clarity.
We impose a \photoz\ quality cut by removing all SB2 and SB3 objects from our bias analysis, 
with $\lesssim10$ per cent loss of objects. 
}
\label{fig:templatetype_photoz}  
\end{figure}

Figure~\ref{fig:templatetype_photoz} shows the photometric redshift
error for the source catalog, for each 
template type separately.  The errors are displayed as a function of \photoz, i.e., the observable;
this format also clearly displays the different features in \photoz\
errors by template types.  The dispersion around the median bias is quantified as
\begin{equation}
\sigma^2_{\Delta z} = \langle [\Delta z - {\rm median}(\Delta z)]^2\rangle \; ,
\end{equation}
and each panel shows $\sigma_{\Delta z}$ for that template type.
The \photoz\ failures\footnote{Photo-$z$'s at the ZEBRA prior limits, 
$\zphot=0$ and $1.5$, are considered \photoz\ failures.} constitute 
$\sim$2 per cent of the objects, which are not shown.
The relative fraction of template types in the \photoz\ sample show that
approximately a quarter each are of the Ell, Sbc, and Scd galaxy
types; about 10 per cent are of the Im-type, and $<$10 per cent are classified as
either SB2 or SB3.  The lens sample shows a similar distribution of galaxy types.

The galaxies classified as starburst (SB2 and SB3) SED types show the largest \photoz\ errors.
Our \photoz\ quality cut discards the SB2 and SB3 template type objects
along with the $\sim$2 per cent \photoz\ failures.
We lose $\lesssim$10 per cent of our sample with this cut.

For the remaining four template types, the width of the \photoz\ error (dashed lines, 
Fig.~\ref{fig:templatetype_photoz}) in general becomes narrower as 
the galaxy becomes brighter, down to $\sigma_{\Delta z}=0.09$ for $r<20.5$.
In general, it is believed that red galaxies (Ell) have better photometric redshift
accuracies compared to the bluer galaxy types, due to the high contrast below and
above the (rest-frame) 4000\AA\ break.  However, this statement is only true when there is a 
strong photometric detection in the bands above and below the break.
The large \photoz\ errors in the Ell and Sbc samples at $\zphot < 0.2$ are 
due to noisy photometry;  this broad dispersion stems from the fact that 
\photoz\ depends on the colour difference between the $g$ and the $ri$
bands at low \zphot\ to probe the $4000\AA$ break; but at dim $ri$ magnitudes, 
the $g$ magnitudes for the red Ell and Sbc galaxies are even dimmer, resulting in 
a \photoz\ fit that is highly sensitive to noise.

\begin{figure}
\includegraphics[trim=0mm 10mm 10mm 23mm, clip, width=84mm]{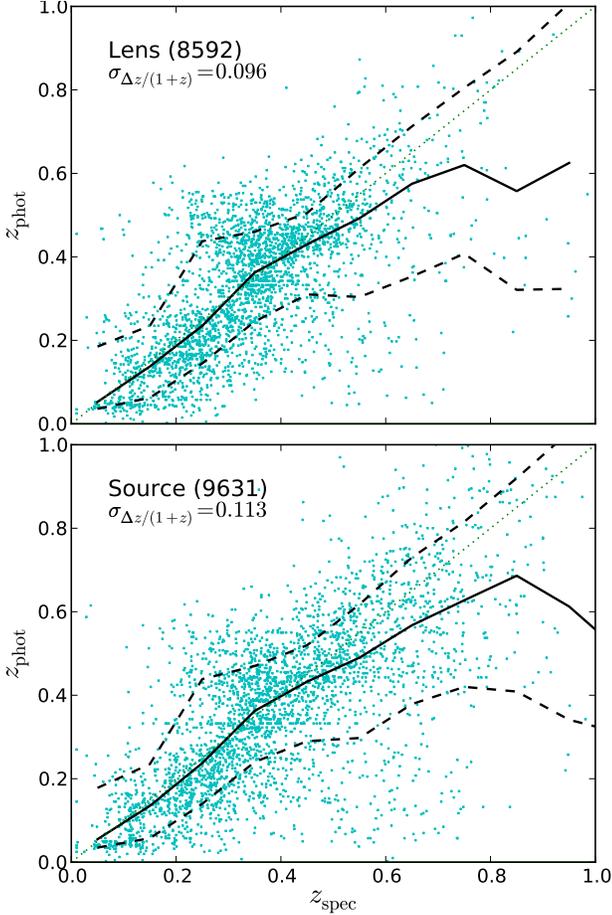}
\caption{
Photo-$z$ as a function of spectroscopic redshifts for the SDSS lens (top) 
and source (bottom) catalogues, after \photoz\ quality cuts have been applied.
The median (bold solid line) and the 16 and 84 percentile range (bold dashed line)
of $\Delta z$ is also shown.
The error uncertainty is given as the standard deviation of $\Delta z/(1+z_{\rm spec})$.  
Only 33 per cent of the points are shown for clarity.
\label{fig:photozspecz}
}
\end{figure}

Figure~\ref{fig:photozspecz} shows results for the full samples (all
templates) using the more standard format for comparison with other
\photoz\ literature: the \photoz's 
are shown as a function of the true redshift, and scatter is quantified as
\begin{equation}
\photozsigma^2 \equiv {\rm var}[\Delta z/(1+\zspec)],
\end{equation}
where $\Delta z\equiv\zphot-\zspec$.
The \photoz\ error is large, where 
$\photozsigma=0.096$ and 0.113 for the lens and the source sample, respectively,
averaged over the whole sample. 
In comparison, ZEBRA template-fitting \photoz's in \citet{feldmann_etal:2006} 
achieve accuracies of $\photozsigma<0.03$
when applied to $I_{AB}<22.5$ COSMOS galaxies with $u^*BVg'r'i'z'K_s$
photometry, thus suggesting that the systematic floor due to
limitations of the code and template set are low, and our errorbars
are dominated by photometric noise.
We find that the scatter is a function of magnitude, and degrades rapidly for $r>21$.  
For our source catalogue, however, the degradation at the faint end is not so severe 
($\photozsigma\sim0.13$ for $21<r<21.8$); this implies that the better resolution 
required for the source sample yields more reliable photometry.

The \photoz\ scatter is asymmetric about the median.
In particular, Figure~\ref{fig:photozspecz} shows bimodality in the \photoz\ distribution
for $0.2<\zspec\lesssim0.4$.  This feature is seen in all template types, and
is an artifact of the gap between the $g$ and $r$-bands at $5440$\AA, or equivalently, 
when the 4000\AA\ break transitions from $g$ into $r$ at $z\sim0.36$.  
The gap between the $g$ and $r$ bands causes uncertainty in the photometric redshift 
when $\zspec=0.36$ even with perfect photometry; with high photometric noise, 
the error in the \photoz\ becomes bimodal centered at this redshift.

\subsection{Uncertainty in template types}
\label{ssec:template_uncertainty}

\begin{figure}
\includegraphics[trim=12mm 0mm 15mm 13mm, clip, width=84mm]{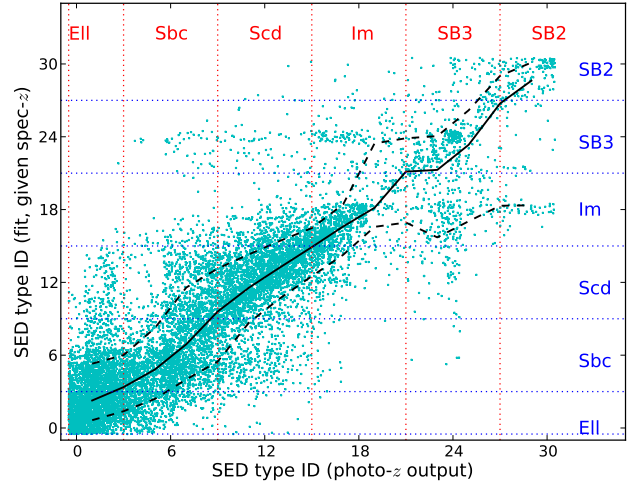}  
\caption{
SED template type uncertainty in the source catalogue (the lens catalogue show similar scatter).
The ordinate is the ``true'' template ID, while the abscissa is the best template type
estimate derived from the \photoz\ run, and is our observable.
The thick solid line and thick dashed lines show the median and the upper/lower
68 percentile scatter, respectively.
The 6 basic CWW+Kinney template types correspond to template type IDs 0, 6, 12, $\ldots$ 30,
with 5 interpolated SED templates in between.
Our SED type classification bins are delineated by the straight, dotted lines. 
(The points are shown with random fractions of $\in[-0.5, 0.5)$ added to better 
illustrate the density; the template type IDs are all integers.)
}
\label{fig:templatetype_uncertainty}  
\end{figure}

It is important to understand the error in template type designation, 
not only because we define our \photoz\ quality cut based upon template type,
but also because we want to understand and minimise the errors in $k$-correction 
(Sec.~\ref{ssec:zebra_kcorrection}) and stellar mass estimates 
(Sec.~\ref{ssec:stellar_mass_estimates}).
Figure~\ref{fig:templatetype_uncertainty} shows the template designation errors
for the source catalogue.
The ``true'' template types were determined by asking ZEBRA to fit to the best template
type, given the known \zspec, and is plotted on the vertical
axis\footnote{The ``true'' template types in this section are true to
  the extent that the ZEBRA template set provides an accurate way to
  classify galaxy templates.  Thus it is meant to represent truth in
  the absence of what photometric noise does to the ability to
  estimate a redshift.};
the horizontal axis is the observable (the template type derived from
photometry alone, when determining the \photoz).
There are 5 interpolated SED templates in between the 6 basic CWW+Kinney template types;
our SED type classification bins are delineated by the straight, dotted lines. 

The median of the scatter lies on the slope $=1$ line, and the scatter width from the median 
in the range of Ell/Sbc/Scd/Im SED types is $\pm$ 3 units,
which is approximately the half-width of the template type bins.
Hence we see that the template type error is reasonably small for the 
chosen template categorisation.
In contrast, very few objects fit to the interpolation between Im to SB2 or SB2 to SB3 types, with 
large uncertainty in the template types in these bins.
The lens catalogue template type uncertainty show similar scatter. 

However, even when using the spectroscopic redshifts, it is possible to have 
the wrong template, either because the template 
set may not describe the true galaxy SED, or due to the relatively large noise 
in the magnitudes which may cause the selection of the wrong template
even when given the true redshift.
This additional template type uncertainty is not accounted for in 
Fig.~\ref{fig:templatetype_uncertainty}.


\section{Bias in Luminosity and Stellar Mass}
\label{sec:application}

To obtain a weak lensing signal with sufficient signal to noise, 
lens galaxies with similar properties (such as luminosity and stellar mass) are stacked.
Absolute magnitudes and stellar mass, used to assign lens galaxies to
each stack, 
inherit scatter from \photoz\ errors.   With our lens calibration set we can measure
this scatter, and use the width of the scatter to determine the
binning resolution for stacking.  Moreover we can test for biases in
the absolute magnitudes and stellar masses, and figure out how to
shift our bins around to account for this type of error.

\subsection{Absolute magnitude}
\label{ssec:absmag_bias}

\begin{figure}
\includegraphics[trim=12mm 0mm 15mm 13mm, clip, width=86mm]{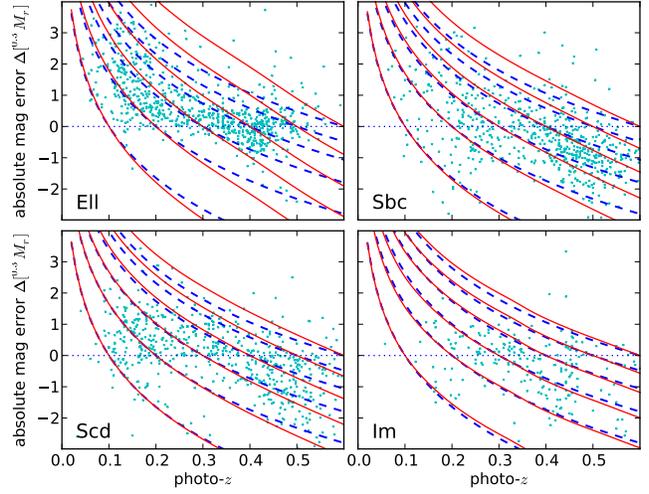}
\caption{
The theoretical $\Mr$ absolute magnitude error curves (distance modulus $+$ $k$-correction)
for objects at \zspec$=0.1, 0.2,$ \ldots, $0.6$ (solid curves, from bottom to top), 
as a function of \zphot, for each template types (Ell, Sbc, Scd, Im).
The dashed line is the distance modulus component alone.  
If the \photoz\ is below the \specz\ (top half of each panel), 
the estimated absolute magnitude is too large (fainter) than the real magnitude.
The correction is more severe at smaller \photoz's.  
The points show distribution of objects from our photometric lens catalogue ($r<21$).
\label{fig:absolutemag_error}  
}
\end{figure}

The bias in absolute luminosity due to \photoz\ errors is a combination of two effects:
the error in the luminosity distance, and in the $k$-correction (Sec.~\ref{ssec:zebra_kcorrection}).
 All magnitudes are $k$-corrected according to SED template types.
We define our absolute magnitude band as the $r$ magnitude redshifted to $z=0.3$ ($\Mr$), 
so that in Eq.~\eqref{eqn:kcorrect}, $R = R' = r$, and $z_0= 0.3$, and where
the observed $r$ magnitude was chosen for its high $S/N$.  
These choices minimise the degree of $k$-correction for our lens sample, and 
most of the correction is dominated by the distance modulus, as demonstrated in
Figure~\ref{fig:absolutemag_error}.  Here,
the theoretical absolute magnitude errors in $\Mr$ (distance modulus $+$ $k$-correction)
are shown in solid curves, for objects at \zspec$=0.1, 0.2,$ \ldots, $0.6$, 
as a function of \zphot.  The dashed lines show the magnitude modification 
from the distance modulus component alone; the fact that they nearly
coincide with the solid lines indicates that the distance modulus
error is more important than the uncertainty in $k$-corrections.  
The points show the distribution of galaxies from our photometric lens catalogue ($r<21$).
Although the $k$-correction is in general small, 
it becomes somewhat more significant for Ell objects at high \zphot\ or \zspec.

\begin{figure}
\includegraphics[trim=12mm 0mm 15mm 13mm, clip, width=86mm]{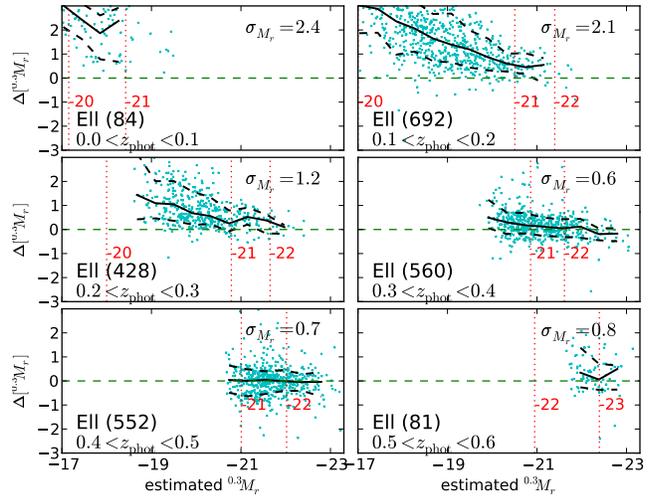}
\caption{
Absolute magnitude bias due to redshift errors, in \zphot\ and $\Mr$ bins, 
for the Ell SED type.  The 6 panels correspond to different \zphot\ bins;
the points show individual objects, while the thick solid and dashed lines 
are the median and 16/84 percentile of the points in a given $\Mr$ bin.
The vertical dotted lines correspond to the true $\Mr$ of the median $\Delta[\Mr]$,
and these lines are labeled with the true $\Mr$ values.
The large bias and scatter at low \photoz\ (\zphot$<0.2$), low luminosity ($\Mr>-21$) 
bins are due to the large \photoz\ errors in those bins for this SED type.
The numbers in parenthesis indicate the galaxy counts in the panel.
\label{fig:absolutemag_error_ell}  
}
\end{figure}

\begin{figure}
\includegraphics[trim=12mm 0mm 15mm 13mm, clip, width=86mm]{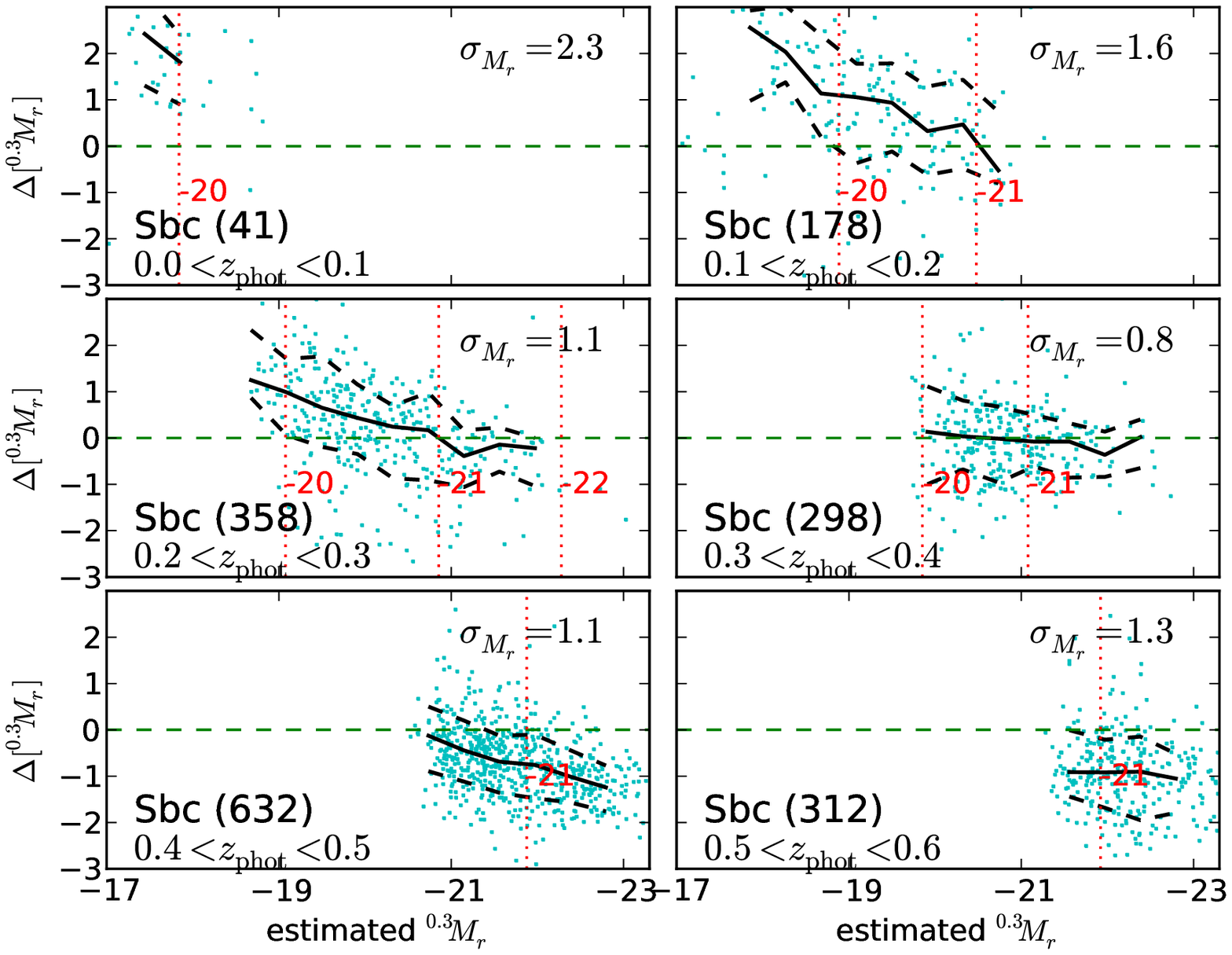}
\caption{
Absolute magnitude bias due to redshift errors, in \zphot\ and $\Mr$ bins, 
for the Sbc SED type (see Fig.~\ref{fig:absolutemag_error_ell} caption
for description).  
\label{fig:absolutemag_error_sbc}  
}
\end{figure}

\begin{figure}
\includegraphics[trim=12mm 0mm 15mm 13mm, clip, width=86mm]{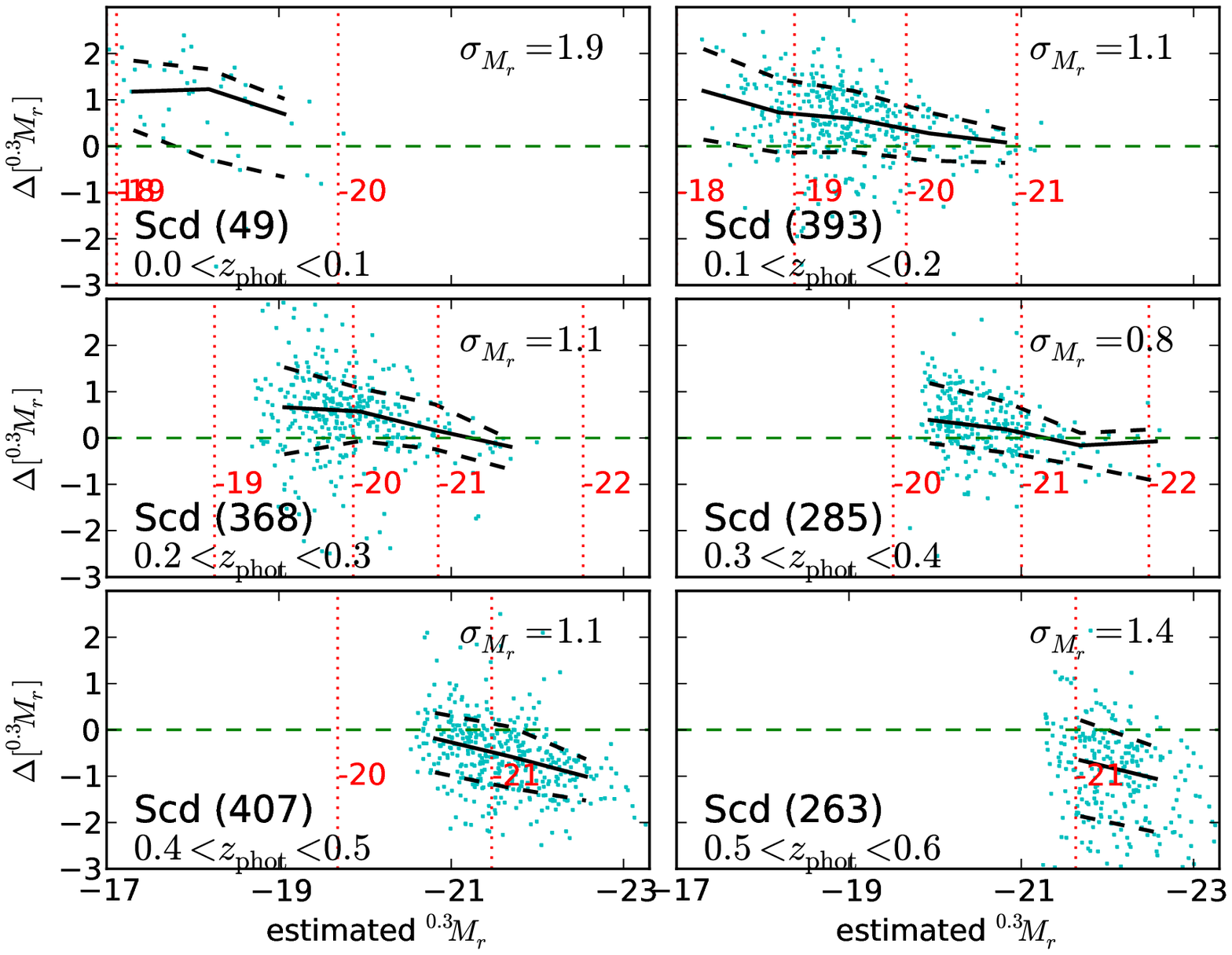}
\caption{
Absolute magnitude bias due to redshift errors, in \zphot\ and $\Mr$ bins, 
for the Scd SED type (see Fig.~\ref{fig:absolutemag_error_ell} caption
for description).
\label{fig:absolutemag_error_scd}  
}
\end{figure}

\begin{figure}
\includegraphics[trim=12mm 0mm 15mm 13mm, clip, width=86mm]{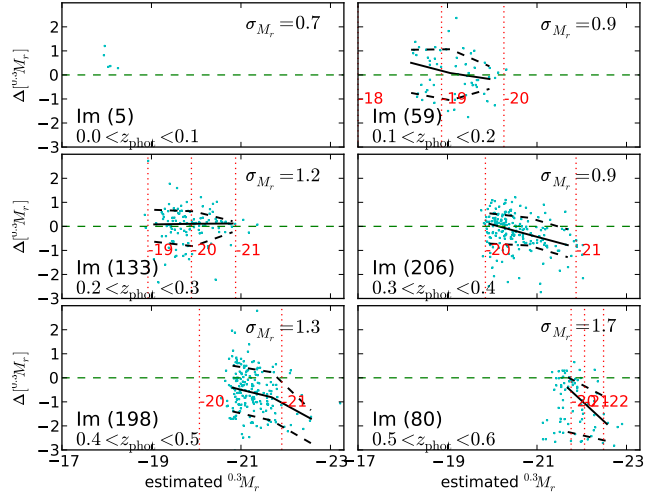}
\caption{
Absolute magnitude bias due to redshift errors, in \zphot\ and $\Mr$ bins, 
for the Im SED type (see Fig.~\ref{fig:absolutemag_error_ell} caption
for description).
\label{fig:absolutemag_error_im}  
}
\end{figure}

The size of the scatter in $\Mr$ dictates the appropriate luminosity binning width.
We further bin the lens objects by redshift bins of width $\Delta\zphot=0.1$, 
up to $\zphot = 0.6$, to determine the binning width as a function of \zphot\ and SED type.  
The actual distribution of absolute magnitude error
$\Delta[\Mr] \equiv\ \Mr(z_{\rm phot}) -\ \Mr(z_{\rm true})$
is determined by the \photoz\ distribution and error distribution of the photometric lens sample.
Figures~\ref{fig:absolutemag_error_ell} through \ref{fig:absolutemag_error_im} show the 
median and scatter in the error $\Delta[\Mr]$ for the four SED
template types $T$ that we use for our analysis, 
as a function of the estimated absolute magnitude $\Mr(z_{\rm phot})$ (the observable),
where sub-panels in each figure are binned by the photometric redshifts.

The smallest scatter is seen in the Ell type galaxies at high redshift ($z>0.3$) and bright
absolute magnitude ($\Mr(z_{\rm phot})<-20$).  This is the binning range most relevant to 
current SDSS LRG studies.  These bright, red galaxies show the smallest 
$\Delta[\Mr]$ scatter of $\sigma_{M_r} \sim 0.7$ magnitude from the median.
The large scatter in the low \zphot, low luminosity region, $\Delta[\Mr]\gtrsim2$,
comes from the broad redshift uncertainty in the red SED types at low $r$ magnitude, 
as discussed in Sec.~\ref{ssec:photozbytemplates}.
Unlike LRGs, dim red objects with $\Mr(\zphot) \sim \Mr(\zspec) > -21$
do not have reliable magnitudes, because the \photoz\ errors are worse (Sec.~\ref{ssec:photozbytemplates}).
For other SED types and \photoz\ bins, a typical magnitude scatter is about 
$\sigma_{M_r}\sim$1, and so the bin size cannot be much smaller than this.  
We also note that where the calibration sample distribution is sparse, 
the binning cannot be reliable.  For this reason, the redshift bin $0<\zphot<0.1$ 
is probably best omitted, for lenses of all SED types.  

The vertical dotted lines in 
Figures~\ref{fig:absolutemag_error_ell}--\ref{fig:absolutemag_error_im} show 
where stacking the lens galaxies by absolute magnitude binning would be meaningful
where the average $\sigma_{\Mr}$ (averaged over the \photoz\ bin) are shown in each panel.
The brightest bins ($\Mr < -21$) are accessible to the Ell and Sbc types at high \zphot,
where the faint end ($\Mr > -19$) are accessible to Scd and Im objects in the 
$0.1<z_{\rm phot}<0.2$ bin.

An additional consideration in comparing magnitudes across different redshifts is
the luminosity evolution.  Between the lowest and highest redshift bin, 
evolution accounts for a difference in $\sim$0.5 magnitudes.
This is estimated according to the luminosity function (LF) evolution,
based on several LF studies in a similar redshift range \citep{wolf_etal:2003,
giallongo_etal:2005, willmer_etal:2006, brown_etal:2007, faber_etal:2007}.
We adopt $Q=1.2$ mag/redshift for all templates, which appears to be consistent with all of 
these studies, and where $Q$ is the redshift evolution slope as defined in
\begin{equation}
M_*(z_0) = M_*(z) + (z-z_0)Q
\end{equation}
where $M_*$ is the magnitude corresponding to $L_*$ in the luminosity function,
$z$ is the object redshift, and $z_0$ is the standard redshift to which we normalise 
our magnitude binning.  
These studies are based on $B$-band luminosities, but $^{0.3}\!r$ and $B$ are close enough
(they partially overlap at 4300--4700\AA); hence the correction from the colour evolution 
is expected to be small.  If we make the simplifying assumption that 
the luminosity rank ordering does not mix as the luminosity function evolves, then we can 
define magnitude bins that shift according to the slope $Q$ as a rough approximation
to the luminosity evolution.  
This correction is a small effect compared to the shifts induced by \photoz\ error bias.

\subsection{Stellar mass}
\label{ssec:stellar_mass}

\subsubsection{Estimating stellar mass}
\label{ssec:stellar_mass_estimates}

Stellar mass can be estimated from the combination of 
(1)  rest-frame colours and (2) absolute magnitude (discussed above in Sec.~\ref{ssec:absmag_bias}),
using correlations by \citet{bell_etal:2003} (via an intermediate step of
estimating a stellar mass-to-light ratio, $\MLR$).  The colour---$\MLR$
relation is defined using the rest frame colour in \citet{bell_etal:2003};
and SED types are used to convert colours and magnitudes into the rest frame.
As excessive $k$-corrections are susceptible to errors in \photoz\ and SED types,
we choose the optimal observed band which minimises the $k$-correction to the rest frame band.
We $k$-correct the observed $i$ magnitude into 
the $^{0.0}\!r$ band rest-frame absolute magnitude 
and  the observed $r-i$ colour to $^{0.0}(g-r)$ 
(Sec.~\ref{ssec:zebra_kcorrection}).  
The $^{0.0}\!r$ band rest-frame is chosen since its wavelength range 
approximately corresponds to that of the observed $i$ band at the lens median redshift 
of $z_0=0.3$, which is the reddest band available in SDSS with high photometric S/N 
(red luminosities require less correction to the $\MLR$ ratio).
Similarly the $^{0.0}(g-r)$ colour is chosen for the higher S/N in the
observed $r$ and $i$ bands.
Then, the $\MLR$ ratio is obtained from Table~7 of \citet{bell_etal:2003},
\begin{equation}
\log_{10}(M_*/L_r) = -0.306 + 1.097 \; [^{0.0}(g-r)]
\label{eqn:stellar_mlratio}
\end{equation}
where $M_*/L_r$ (stellar mass $M_*$ to $^{0.0}r$-band luminosity ratio) is in solar units.
The $r$ band luminosity in solar units is
\begin{equation}
\log_{10}L_r = -0.4 \; (^{0.0}\!M_r - 4.64)
\label{eqn:lum_mag_relation}
\end{equation}
where $(M_r)_\odot = 4.64$ is the $r$-band absolute AB magnitude of the Sun 
\citep{blanton_roweis:2007}.  
Hence the stellar mass in solar units is 
\begin{equation}
\log_{10}M_*  =  1.550 + 1.097 \; [^{0.0}(g-r)] - 0.4 \; [^{0.0}\!M_r].
\label{eqn:stellar_mass}
\end{equation}
These numbers quoted above are based on a diet Salpeter IMF; for future science work 
we can modify for this by rescaling all stellar masses by a constant factor.

We now consider biases in these stellar masses.  We start with biases in the
intermediate step of constructing the stellar mass-to-light ratio.

\subsubsection{Bias in stellar mass-to-light ratio}

\begin{figure}
\includegraphics[trim=5mm 0mm 15mm 13mm, clip, width=86mm]{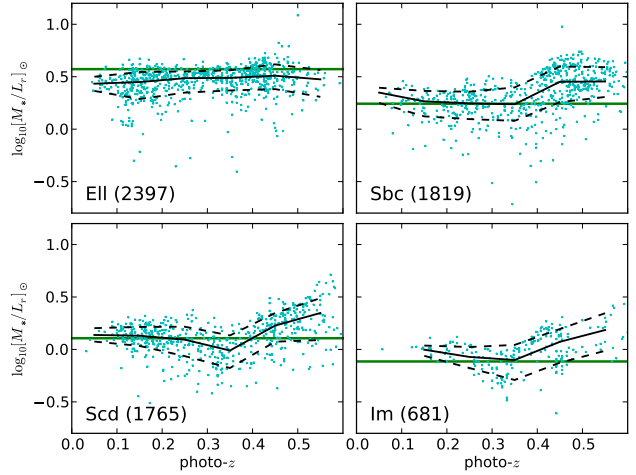}
\caption{
The scatter in the stellar mass-to-light ratio ($\MLR$) 
with respect to rest-frame $r$-band luminosity in solar units, 
as a function of \photoz, for the different SED types, determined from \photoz\ fitting.
The thick horizontal line shows the theoretical $\MLR$ based on the rest-frame colour,
and is constant for a given SED type.
The points are the $\MLR$ estimated from the ``true'' template type 
(best-fit template given the \specz) $k$-corrected using the true redshift (\zspec), 
and the solid/dashed lines show the median and 68 percentile scatter, respectively.
The $\MLR$ determined from \zphot\ has intrinsic scatter of $\sim$0.1 dex (not shown).  
The scatter in $\MLR$ is small compared to the bias in the luminosity $L$.
\label{fig:stellar_mlratio}  
}
\end{figure}

In this section, we consider errors in the inferred $\MLR$ using
\photoz, starting from the basic assumption that these are
significantly larger than intrinsic uncertainties in the
\citet{bell_etal:2003} prescription for estimating stellar mass when
using spectroscopic redshifts.  This statement ignores underlying
uncertainties in the stellar IMF, which plague all stellar mass
estimates.
 
There is very little bias and scatter when we plot the $\MLR$ obtained from the rest-frame 
$^{0.0}(g-r)$ colour by SED template types.  This is because for a 
given SED template, the rest-frame colour is a constant.
Instead, we take the ``true''  template type\footnote{The ``true'' template type is 
the best-fit SED template given \zspec, which is unbiased 
but scattered with respect to the observed template type; 
see Fig.~\ref{fig:templatetype_uncertainty}.} and $k$-correct using the true redshift.
The $k$-corrected colour $^{0.0}(g-r)$, and hence the derived $\MLR$, then show some 
bias and scatter, as seen in Figure~\ref{fig:stellar_mlratio}.
The constant (as a function of \zphot) predicted $\MLR$ for the given template type 
is shown for comparison.
The points show the distribution of the lens catalogue objects, 
and the solid/dashed lines show the median and 68 percentile scatter, respectively.
There is still little variation in the $\MLR$ evident, because
the rest-frame colours are identical for a given template type, and the 
``true'' template type deviates little from the estimated one.
The magnitude of the scatter shown is similar to the uncertainty in the $\MLR$ relation
(Eq.~\ref{eqn:stellar_mlratio}) itself, which is $\sim0.1$ dex \citep{bell_etal:2003}.
The true errors in $\MLR$ might be expected to be larger than that shown here, given that 
the actual galaxy SED is unknown.  However, we expect the error in the luminosity to 
dominate, discussed below.

\subsubsection{Bias in stellar mass}

\begin{figure}
\includegraphics[trim=6mm 0mm 15mm 10mm, clip, width=86mm]{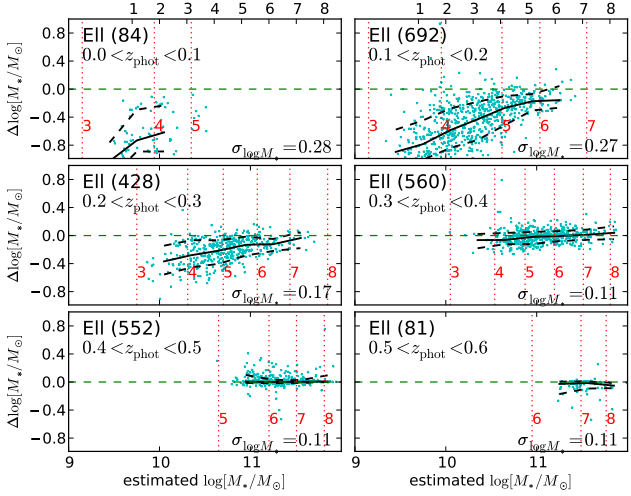}
\caption{
Stellar mass errors $\Delta\log_{10}[M_*/M_\odot] \equiv
\log_{10}[M_*/M_\odot](z_{\rm phot}) - \log_{10}[M_*/M_\odot](z_{\rm spec})$
in the Ell SED type, as a function of the estimated stellar mass based on \zphot\ 
(the observable).  
The distribution of the lens catalogue are plotted as points, where the median and 
68 percentile scatter are shown as the thick solid and dashed lines, respectively.
Each panel correspond to different \zphot\ bins.
The top axis of the top panels show the nominal stellar mass bin of 0.3 dex, or where
the bin is twice as massive as the previous one.
The vertical dotted lines show the bias-corrected binning.
\label{fig:stellar_mass_ell}  
}
\end{figure}

\begin{figure}
\includegraphics[trim=6mm 0mm 15mm 10mm, clip, width=86mm]{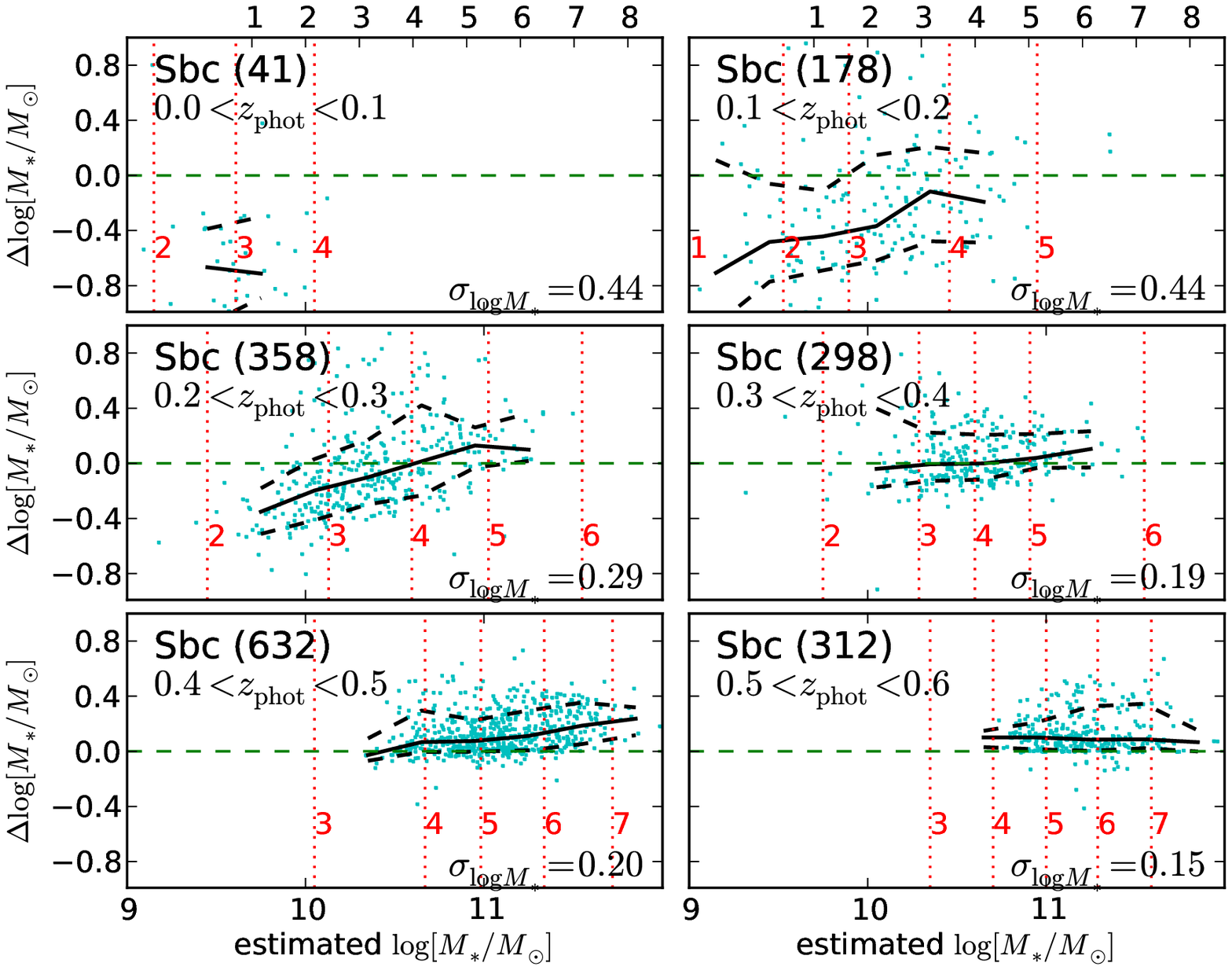}
\caption{
Stellar mass errors in the Sbc SED type.  
\label{fig:stellar_mass_sbc}  
}
\end{figure}

\begin{figure}
\includegraphics[trim=6mm 0mm 15mm 10mm, clip, width=86mm]{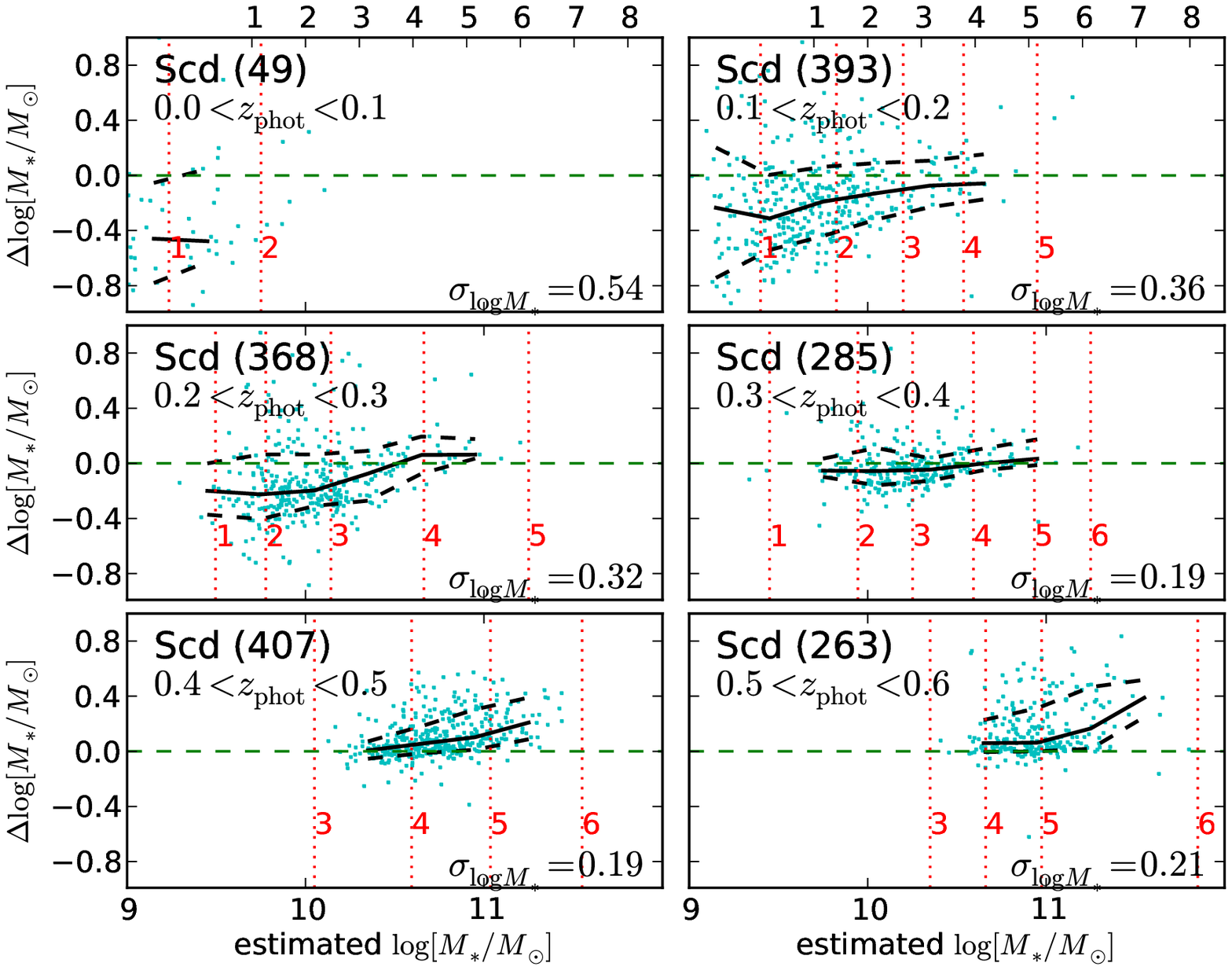}
\caption{
Stellar mass errors in the Scd SED type.  
\label{fig:stellar_mass_scd}  
}
\end{figure}

\begin{figure}
\includegraphics[trim=6mm 0mm 15mm 10mm, clip, width=86mm]{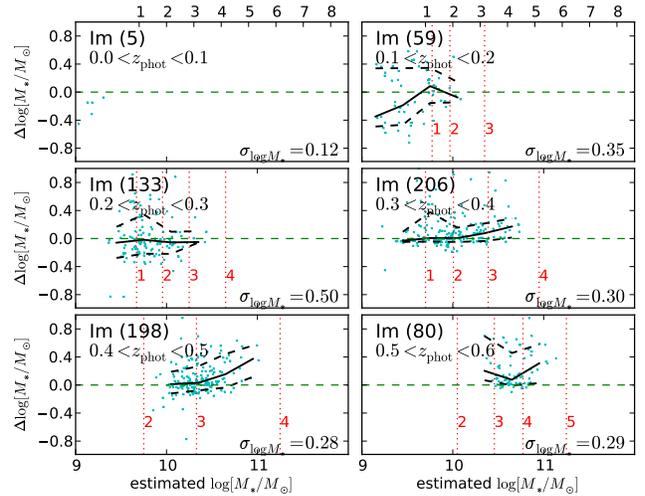}
\caption{
Stellar mass errors in the Im SED type.  
\label{fig:stellar_mass_im}  
}
\end{figure}

While the scatter in $\MLR$ is of order $0.1\sim0.2$ dex, the scatter in the luminosity 
is $\sim0.4$ dex (converted from $\sim$1 mag scatter, see Eq.~\ref{eqn:lum_mag_relation}),
mostly dominated by the distance modulus (Sec.~\ref{ssec:absmag_bias}).
Figures~\ref{fig:stellar_mass_ell}--\ref{fig:stellar_mass_im} show the bias and 
scatter in stellar mass estimate in the four template types, 
as a function of estimated stellar mass, for the 6 \photoz\ bins.
The errors (vertical axis) are plotted as a function of estimated stellar magnitudes
$\log_{10}[M_*/M_\odot](z_{\rm phot})$ (the observable), where the nominal stellar mass bins 
are in widths of 0.3 dex (i.e, each bin is twice as massive as the previous bin). 
These bins are labeled by the index 1 through 8, as seen on the topmost panels in each Figure.
The vertical dotted line, labeled by stellar mass bin numbers, are the corrected stellar masses
based on the median values.

We see the general trend in the absolute luminosity bias and scatter repeated in these plots, e.g.,
the Ell types show small scatter at high stellar mass and high \zphot, 
and the lowest \zphot\ bin ($0<\zphot<0.1$) is too sparse.
The $\log_{10}[M_*/M_\odot]$ scatter $\sigma_{\log M_*}$, which is defined with respect 
to the median, is shown in each plot, and indicates small scatter for most bins.
In fact, most bins show scatter smaller than the nominal 0.3 dex stellar mass bin.

We note that previous studies (e.g., \cite{dutton_etal:2010}) have
compared the photometry-based stellar mass estimate employed here 
with spectra-based estimates.
On average, the two agree for higher stellar masses, but
deviate systematically at lower stellar mass.  Such systematic deviation can be corrected 
accordingly with a stellar-mass dependent correction.

\section{Bias in the gravitational lensing signal}
\label{sec:calib_lens}

With these photometric redshifts, their scatter and their effects on stellar mass and 
absolute magnitudes quantified, we can turn to the bias and scatter induced in the 
gravitational lensing signal due to \photoz\ use for sources and lenses.
As the conversion of lensing shear to mass depends upon the redshifts of the 
lens and source (see Eq.~\ref{E:sigmacrit}), errors in these redshifts 
propagate into errors in the weak lensing measurements.  
We consider three cases, 
(a) the lens redshift is known and \photoz\ is only used for the source,
(b) the source redshift is known and \photoz\ is only used for the lens, and
(c) \photoz's are used for both lens and source.
We deal with these in turn below, focusing on biases in the lensing
signal calibration; in principle, there is also a blurring of
information in the transverse direction once \photoz\ are used for
lenses; however, a typical 10 per cent \photoz\ error does not have a
significant effect on the shape of the lensing profiles, so we neglect
this effect here.  We begin by reviewing and then
expanding upon 
the methods by \citet{mandelbaum_etal:2008} to calculate lensing bias.  

\subsection{Methods}
\subsubsection{Bias}
In the absence of \photoz\ errors, the observed lensing tangential
shear $\gamma_t$ can be related to the lens surface density contrast 
$\Delta\Sigma$ via
\begin{equation}
\Delta\Sigma =  \Sigma_c \gamma_t
\end{equation}
where the proportionality constant, the critical surface density, was
defined in Eq.~\eqref{E:sigmacrit}. 


The redshift calibration bias is the misestimation of $\Delta\Sigma$ due to the \photoz\ scatter.
To estimate this bias, we first consider the method of estimating
$\Delta\Sigma$, via a weight-average over the tangential component of the 
measured source galaxy shapes $\tilde\gamma_t^{(j)}$,
\begin{equation}
\widetilde{\Delta\Sigma} = 
  \frac{\sum_j \tilde{w}_j \tilde\gamma_t^{(j)} \widetilde\Sigma_{c,j}}{\sum_j \tilde{w}_j} \; .
\end{equation}
Here, the summation
is over  $j$ lens-source pairs, $\Sigma_{c,j}$ is the critical mass density for the $j$th
lens-source pair, and the tilde
indicates estimated values (using \photoz's where applicable).  The optimal weight is
\begin{equation}
\tilde{w}_j = 
  \frac{1}{\widetilde\Sigma_{c,j}^2 \, (e^2_{\rm rms} + \sigma_{e,j}^2)}.
\label{eqn:weights}
\end{equation}
The quantities added in quadrature are
$e_{\rm rms}$, the shape noise of the source ensemble, and $\sigma_{e,j}$, the shape 
measurement error of the $j$th galaxy (per single component of the shear).
Assuming that the only calibration bias occurs through the use of \photoz\ via the quantity $\widetilde\Sigma_c$, 
the redshift calibration bias $b_z$ is
the ratio of $\widetilde{\Delta\Sigma}$ to the true signal
$\Delta\Sigma$.  By substituting $\tilde\gamma_t^{(j)} =
\Delta\Sigma\Sigma^{-1}_{c,j}$, we find
\begin{equation}
b_z(\zlens) + 1 \equiv \frac{\widetilde{\Delta\Sigma}}{\Delta\Sigma}
 = \frac{\sum_j \tilde{w}_j \Sigma^{-1}_{c,j}\widetilde\Sigma_{c,j}}{\sum_j \tilde{w}_j} \; ,
\label{eq:single_bias}
\end{equation}
the weighted sum of the ratio of the estimated to true critical
surface density.

In the actual bias estimation,
each galaxy $j$ is further weighted to
``smooth'' out the LSS in the calibration sample.  To do so, we fit
the redshift histogram to the 
analytic curve Eq.~\eqref{eqn:fitting_formula}, as seen in 
Figs.~\ref{fig:source_specz_distribution} and \ref{fig:lens_specz_distribution};
the LSS weight for all galaxies in a particular histogram bin $i$ is then 
the ratio of the number of galaxies according to the fit 
distribution, to the real number of galaxies, i.e.,
$w_{\rm LSS} = N^{\rm (model)}_i / N_i$ \citep{mandelbaum_etal:2008}.  
The analytic curve is fit separately for every calibration subsample that is used.

The bias (Eq.~\ref{eq:single_bias}) is for a single lens redshift.
If we want to estimate the average bias from using \photoz\ for \zsrc\
over all lenses with known \zlens, we need to average over the lens
redshift distribution, 
including source weight factors,
\begin{equation}
\langle b_z \rangle 
= \frac{\int \rmd\zlens\,p(\zlens)\tilde{w}_l(\zlens) b_z(\zlens)}
{\int \rmd\zlens\,p(\zlens)\tilde{w_l}(\zlens)}  \; .
\end{equation}
The weight for a given lens redshift $\tilde{w}_l(\zlens)$ is 
\begin{equation}
\tilde{w}_l(\zlens) = 
  D^{-2}_L (1+\zlens)^{-2} \, \sum_k \tilde{w}_k \; .
\label{eqn:area_weight}
\end{equation}
Here the summation is over all source galaxies $k$ estimated to be beyond the lens redshift $\zlens$, and
$\zlens$ may be spectroscopic or photometric.  The reason for the
prefactors before the summation is that our method of estimating these
sums using a calibration sample of fixed area does not really
correspond to how lensing signals are actually measured.  Typically
they are estimated within some fixed physical or comoving aperture,
which means that lenses at lower redshift will use sources from an
effectively larger area on the sky, and therefore get greater weight
according to the square of that angular diameter distance to the lens
redshift.  The factor of $(1+\zlens)^{-2}$ assumes the use of a fixed
comoving aperture.  Note that this effect was incorrectly neglected in the previous analysis 
(eq.~6 of \citealt{mandelbaum_etal:2008}).  Fortunately, it is only
significant when averaging over a broad lens redshift distribution.
In subsequent papers relying on the \cite{mandelbaum_etal:2008}
lensing signal calibrations, including \cite{mandelbaum_etal:2008b,reyes_etal:2008,
mandelbaum_etal:2009, mandelbaum_etal:2010, schulz_etal:2010}, the change in the
lensing signal calibration when including this additional redshift
weighting factor is typically $2$ per cent, which is approximately the
size of the quoted systematic uncertainty in the redshift calibration,
and typically ~20 per cent of the statistical error.   



\subsubsection{Source-lens pair (statistical) incompleteness}
\label{ssec:purity_and_completeness}

The bias as described above does not indicate the statistical loss 
in the number of valid source-lens pairs, or in the deviation from
optimal weighting due to the use of \photoz.
Such contributions can be described in terms of the purity and completeness of the 
lens-source pairs, and the change in weights from using \photoz\ instead of the true redshift.
Purity is the fraction of lens-source pairs that are truly lensed with all ``valid'' pairs based on \photoz,
where each pair is weighted according to Eq.~\eqref{eqn:weights}.
Completeness is the ratio of the sum of all weights of the lens-source pair ensemble, 
calculated based on \photoz\ to that based on the true redshift.
The loss of statistics due to low purity and completeness translates to
higher uncertainty (variance) in $\Delta\Sigma$.
The efficiency is defined as the ratio of the variances, where the smaller, ideal variance is attained
when all redshifts are known, and the estimated variance obtained from using \photoz's
is always larger than the ideal variance due to the two previous
issues (low purity and/or completeness) plus deviation of the
weighting from the optimal weighting.  This efficiency is given by 
\citep{mandelbaum_etal:2008}
\begin{equation}
\frac{{\rm Ideal\ Var}(\Delta\Sigma)}{{\rm Real\ Var}(\Delta\Sigma)} =
\frac{ \left( \sum\sqrt{\tilde{w}w}\right)^2}{\left(\sum{w}\right)\left(\sum{\tilde{w}}\right)}
\end{equation}
where the weights are defined in Eq.~\eqref{eqn:weights}, and the tilde indicates 
values estimated based on \photoz's.
These will be illustrated for the case of both source and lens \photoz\
(\S \ref{ssec:both_lensbias}) below.  

\subsubsection{Bias uncertainties from LSS}
\label{ssec:lss_bias_uncertainties}

The uncertainty in the calibration bias estimated using
Eq.~\eqref{eq:single_bias} is determined not only by the size of the
calibration sample (Poisson statistics) but also by its LSS
fluctuations.    
To estimate the size of this uncertainty, \citet{mandelbaum_etal:2008} used a modified bootstrap resampling
method to estimate the uncertainty in the true source redshift
distribution, and hence in the 
bias curve, $b_z(\zlens)$, where \zlens\ is known.  
In this paper, we take a simpler approach.  Since we have several calibration subsamples
representing independent realizations of LSS, we obtain a bias curve
$b_z(\zlens)$ for each subsample, 
take their average as the bias, and the standard deviation of the mean
as the systematic uncertainty due to LSS (see, e.g.
Fig. ~\ref{fig:sourcephotozbias}).  

Given the formalism described in this section, we now turn to the case
of source \photoz, lens \photoz, and both combined.

\subsection{Lensing bias in source \photoz}
\label{ssec:source_lensbias}

\begin{figure}
\includegraphics[trim=0mm 0mm 15mm 11mm, clip, width=84mm]{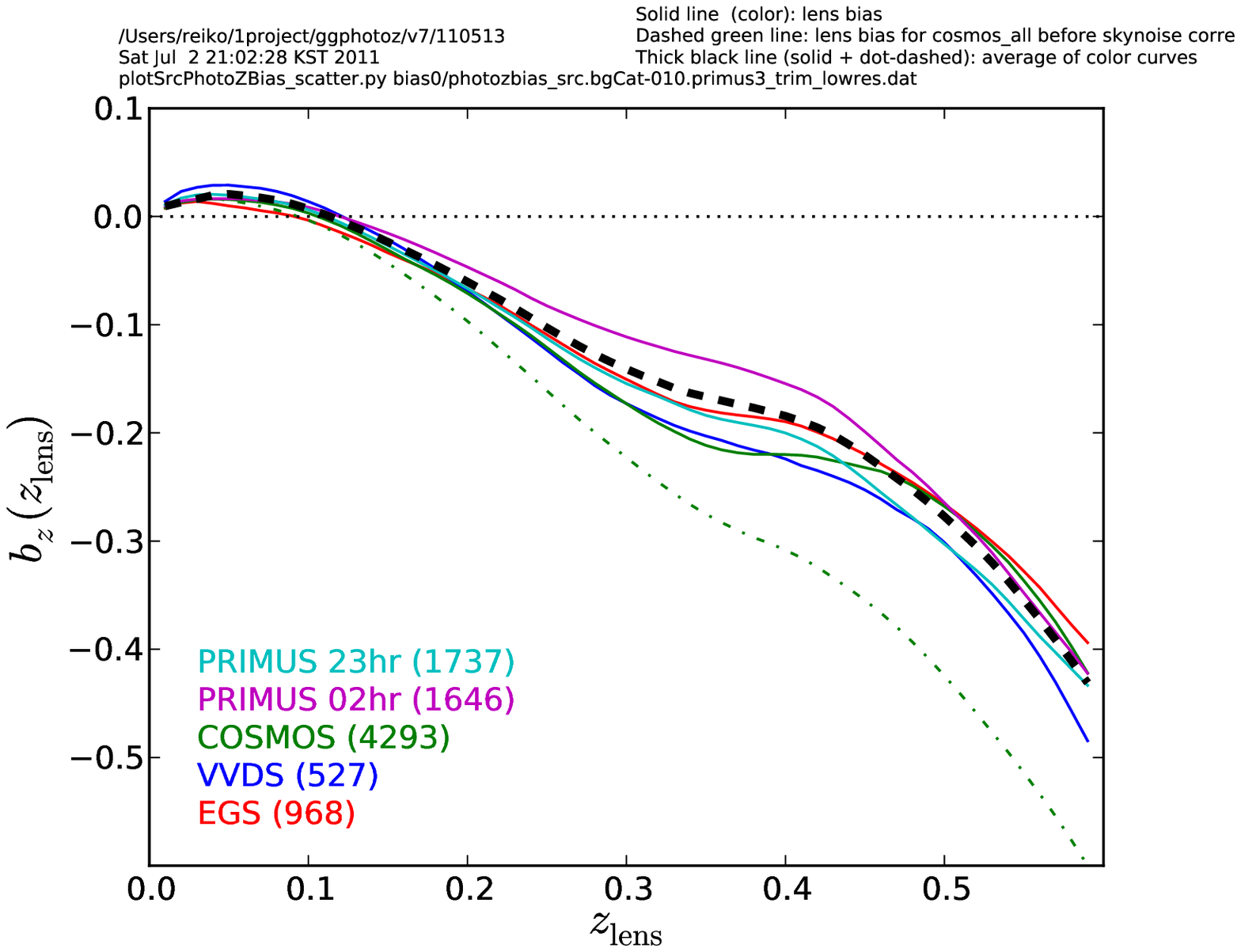}
\caption{
Lensing signal calibration bias from using \photoz\ for \zsrc, given a known lens redshift \zlens.
The thin solid lines show bias in the EGS, VVDS, COSMOS, PRIMUS-02hr and PRIMUS-23hr fields, 
and the weighted average of the 5 solid curves is shown as the thick dashed line.
Of the 5 fields, the COSMOS field bias has been empirically corrected (see text); 
the pre-correction bias is shown as the thin dot-dashed line.  
\label{fig:sourcephotozbias}  
}
\end{figure}

First, we consider the case where the lens redshifts are known to high accuracy, but
the source redshifts rely on much more uncertain \photoz.  
Figure~\ref{fig:sourcephotozbias} shows the lensing bias $b_z(\zlens)$ for using \photoz\ in
the new SDSS source catalogue as a function of the known lens redshifts \zlens.
Bias for each the 5 source calibration subsamples (EGS, VVDS, COSMOS, PRIMUS-02hr and PRIMUS-23hr) are shown as thin solid lines.
Of the 5 fields, the calibration bias in the COSMOS field has been
empirically corrected to account for excessive sky noise, as will be
described shortly.
Prior to calculating the bias, all subsamples have 
the \photoz\ quality cuts applied (Sec.~\ref{ssec:photozbytemplates}),
and the ``excess'' low-resolution galaxies have been omitted
in the VVDS, PRIMUS-02hr and 23hr fields (Sec.~\ref{ssec:obs_correction}).
The bias $b_z(\zlens)$ can be as low as $-0.4$ at $\zlens=0.5$, but the uncertainty in the
bias is less than 0.03; i.e., $-0.40\pm0.03$.
In general, source \photoz's with large scatter tend to pull the bias
more negative. 

As noted previously, it was impossible to augment the COSMOS sample to
correct for the loss of observable galaxies due to excessively high
sky noise (Sec.~\ref{ssec:sdss_obs_conditions}); instead, it is
necessary to derive an empirical correction to the bias.  
The redshift calibration bias curve for the COSMOS calibration
sample, before correction, can be seen as the dot-dashed line in
Fig.~\ref{fig:sourcephotozbias}; clearly it is quite discrepant
compared to the other samples.  
We have empirically compensated for this effect using the following procedure:
We used the fact that our PRIMUS-02hr and 23hr calibration samples are
on stripe 82, where there are many SDSS observing runs with different
conditions.  Thus, we chose runs overlapping those two regions with
seeing that is comparable to in the COSMOS region, but with two
different values of sky noise: typical, and 20 per cent larger (as in
the COSMOS region).  
We carry out the bias calculations independently for each of the
observing runs with different conditions, including matching against
the spectroscopic training sample, imposing the lensing quality cuts,
and obtaining
\photoz's from the photometry.  We calculate the resulting lensing
signal bias $b_z(\zlens)$ from the \photoz's for the two runs separately,
and then use the difference between to two to correct for the COSMOS
sample bias.  It is reassuring that after applying this correction,
the calibration bias for the COSMOS sample is reasonably consistent
with that for the other four calibration samples.

\subsection{Lensing bias in lens \photoz}
\label{ssec:lens_lensbias}

\begin{figure}
\includegraphics[trim=1mm 10mm 15mm 20mm, clip, width=84mm]{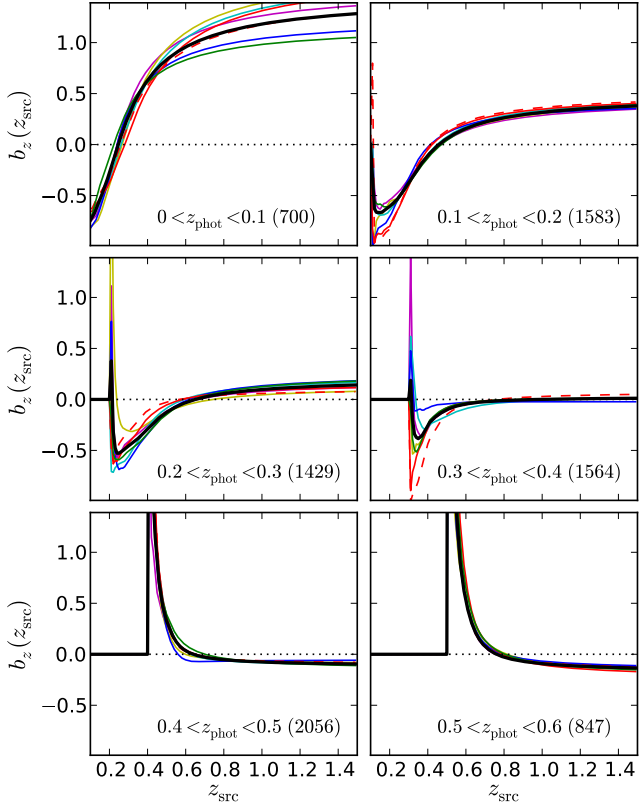}
\caption{
Bias in the gravitational lensing signal due to \photoz\ usage for lens redshifts,
as a function of known source redshift, $b_z(\zsrc)$.
The different panels show \zphot\ bins of lens redshifts; the number in parenthesis
indicates the number counts of galaxies in the spectroscopic lens
calibration sample in this \photoz\ bin.
The thin lines are the biases in the six individual lens calibration subsamples,
and the thick line is the overall bias from the combination of all
calibration subsamples. The thin dashed line indicates the EGS sample before the 
deficient $z\sim0.3$ bin has been corrected for (Sec.~\ref{ssec:lens_lensbias}).
We discard the lowest redshift bin, $0<z_{\rm phot}<0.1$, since the different
calibration subsamples suggest inconsistent values for the calibration bias.
\label{fig:lensphotozbias}  
}
\end{figure}

Next we calculate the bias in the gravitational lensing signal when the source
redshift is known, but \photoz\ is used for the lens redshift.
Unlike the case above (Sec.~\ref{ssec:source_lensbias}) where \zlens\ is known but
\photoz\ is used for source redshifts, such a scenario is not relevant for 
real observations.  However, this exercise will allow us to study the general response
of the lensing bias $b_z$ to errors in lens \photoz, and may suggest
certain scenarios in which using \photoz\ for \zlens\ 
would be inadvisable.  As this is just an exercise, we do not
consider the purity, completeness and efficiency; only the lensing
calibration bias is shown.

Figure~\ref{fig:lensphotozbias} shows the bias $b_z(\zsrc)$ as a function of the 
known \zsrc.  We initially found that all calibration subsamples agree well except for EGS;
this deviation was due to the lack of objects in the $z\sim0.3$ bin, as
discussed in Sec.~\ref{ssec:deficient_redshift_bins}.  This discrepancy is easily 
remedied when the deficient bin in the EGS sample is replaced with, e.g., 
objects from the VVDS survey of the same redshift bin (both the original and 
corrected EGS biases are shown in the plot).  Note that this correction was not required in
Sec.~\ref{ssec:source_lensbias} when using source \photoz.  
The observing conditions that affect the $r<21.8$ source calibration sample are
not relevant to the lens $r<21$ sample because (a) it does not have resolution
cuts, and hence is not affected by the variation in the seeing, and (b) it has 
a brighter flux limit, such that the sky noise has minimal effect on the number counts
(Sec.~\ref{ssec:obs_cond_effect}).

For all \zlens\ bins, we see that the lensing bias converges at high \zsrc\ 
to a reasonable value for most \zphot\ bins except the lowest lens redshift bin
$0<z_{\rm phot}<0.1$.
We also see that the bias $b_z(\zsrc)$ varies rapidly close to the lens redshift,
and also shows the largest difference between the calibration subsamples
(perhaps reflecting a sensitivity to LSS fluctuations in the lens
number density or overall characteristics such as domination by red
vs. blue galaxies).
The characteristic width of this high-variation region is approximately 
$\Delta\zphot\sim0.1$, implying that a minimum lens-source separation of 0.1 
in redshift can help stabilise the bias and minimize its associated
systematic uncertainty.  In reality, with only \photoz\ 
for both \zlens\ and \zsrc, such a clean cut is impossible,
and there may not be significant benefit when applying a minimum separation cut.
We explore this issue further in the next section.

\subsection{Lensing bias, lens and source \photoz}
\label{ssec:both_lensbias}

\begin{figure}
\includegraphics[trim=9mm 5mm 19mm 13mm, clip, width=84mm]{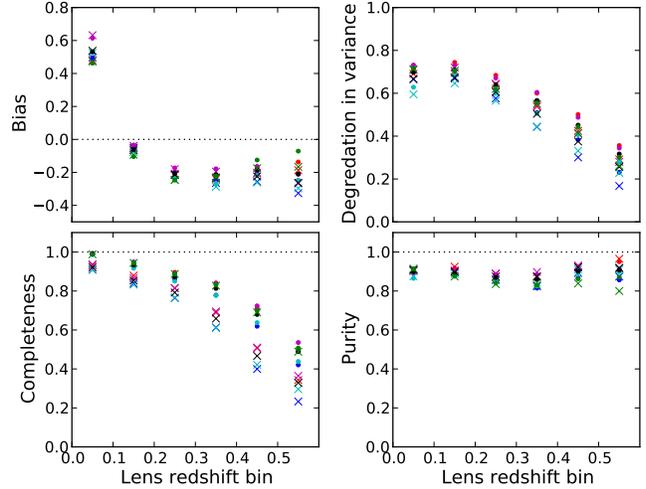}
\caption{
{\em Top left:}
Bias $b_z$ in the galaxy-galaxy lensing signal due to \photoz\ usage for both the 
lens and source.  The lens galaxies have been binned according to \photoz\ bin
width of $\Delta\zphot=0.1$.
All 5 calibration subsamples (EGS, VVDS, COSMOS, PRIMUS-02hr and PRIMUS-23hr) 
have been plotted to show the LSS-induced scatter.
Dots: no pair rejection applied.  Crosses: Minimum \photoz\ separation of 0.1 
between lens and source required.  
{\em Top right:}  
Efficiency (degradation in variance), as defined in Sec.~\ref{ssec:purity_and_completeness}.
{\em Bottom left and right panels:} Completeness and purity, respectively,
as defined in Sec.~\ref{ssec:purity_and_completeness}.
The weighting scheme keeps the lensing signal relatively pure, while the degradation in 
the variance (statistical loss) comes mostly from the incompleteness of the valid 
lens-source pairs due to the use of \photoz\ errors.
\label{fig:lenssourcephotozbias}  
}
\end{figure}

\begin{table}
\caption{Lensing signal bias and its uncertainty, for \photoz\ lens and source redshifts,
from the bias panel in Figure~\ref{fig:lenssourcephotozbias}.
}
\label{tab:lens-src_photoz_bias}
\begin{tabular}{c r l}
\hline
Lens \photoz\ bin & Lensing bias & (COSMOS noise bias) \\
\hline
\multicolumn{3}{l}{Using all samples}\\
0.05 &   0.513 $\pm$ 0.055 & \ \ \ \ \ (-0.027) \\
0.15 &  -0.061 $\pm$ 0.025 & \ \ \ \ \ (-0.035) \\
0.25 &  -0.211 $\pm$ 0.027 & \ \ \ \ \ (-0.048) \\
0.35 &  -0.227 $\pm$ 0.029 & \ \ \ \ \ (-0.068) \\
0.45 &  -0.178 $\pm$ 0.034 & \ \ \ \ \ (-0.105) \\
0.55 &  -0.183 $\pm$ 0.069 & \ \ \ \ \ (-0.167) \\
\hline
\multicolumn{3}{l}{With minimum \photoz\ separation of 0.1}\\
0.05 &  0.521 $\pm$ 0.058 & \ \ \ \ \ (-0.027) \\
0.15 & -0.063 $\pm$ 0.019 & \ \ \ \ \ (-0.041) \\
0.25 & -0.219 $\pm$ 0.025 & \ \ \ \ \ (-0.049) \\
0.35 & -0.244 $\pm$ 0.031 & \ \ \ \ \ (-0.074) \\
0.45 & -0.220 $\pm$ 0.033 & \ \ \ \ \ (-0.115) \\
0.55 & -0.249 $\pm$ 0.062 & \ \ \ \ \ (-0.167) \\
\hline
\end{tabular}
\end{table}

One of the main goals of this paper is to lay the groundwork for galaxy-galaxy
lensing using \photoz\ for sources and lenses.  As mentioned
in the introduction, this will dramatically increase the range of SDSS
galaxies accessible for study with galaxy-galaxy lensing, both in the
direction of lower luminosity at fixed redshift, and also going to
higher redshift samples at fixed luminosity.
Figure~\ref{fig:lenssourcephotozbias} shows the lensing signal bias, efficiency,
completeness and purity when \photoz\ lens-source pairs are used.  
The bias values have been tabulated in Table~\ref{tab:lens-src_photoz_bias}.
The lens galaxies have been binned by \photoz\ only.  
All 5 calibration subsamples (EGS, VVDS, COSMOS, PRIMUS-02hr and 23hr) 
have been plotted to show the LSS-induced scatter.
Each subsample has been modified as follows:
(1) the COSMOS sample has been corrected for the skynoise-induced excess bias
(Sec.~\ref{ssec:source_lensbias}), where the excess bias is tabulated in 
Table~\ref{tab:lens-src_photoz_bias};
(2) the modified EGS lens sample has been used (Sec.~\ref{ssec:lens_lensbias})
to calculate the bias in EGS; and (3) the low resolution objects in the 
stripe 82 samples (VVDS, PRIMUS-02hr and PRIMUS-23hr) have been removed from 
the source sample (Sec.~\ref{ssec:obs_correction}).
Although not shown here, further binning of the lens objects
(reflecting the stacking scheme, e.g. based on stellar mass or
luminosity) would be necessary to estimate the bias for any particular
science application.

These panels again show that calibrating the lowest redshift bin for
the lens can be unreliable, as demonstrated by the large variance in
the bias $b_z$ between the different calibration samples.
The highest redshift bin have a significant scatter in the bias among the subsamples
which comes from the skynoise correction (applied to the COSMOS sample).
Otherwise, the bias correction is $\sim$20 per cent, with the subsamples showing agreement
to 4 per cent scatter.
As we go to higher lens redshift bins, the efficiency and completeness drops, but not too
significantly.  Purity (at 90 per cent) remains high in all lens redshift bins.
Application of lens-source \photoz\ separation requirement does not have a 
significant effect on the lensing bias or its scatter, while completeness 
(and hence efficiency) drops.
This indicates that the minimum \photoz\ separation cut as implemented
here is unnecessary.  We interpret this to mean that
our weighting scheme (Eq.~\ref{eqn:weights}) sufficiently down-weights
nearby lens-source pairs that their potentially strong calibration
biases do not lead to significant uncertainty in the calibration bias.

\begin{figure}
\includegraphics[trim=20mm 30mm 20mm 13mm, clip, width=84mm]{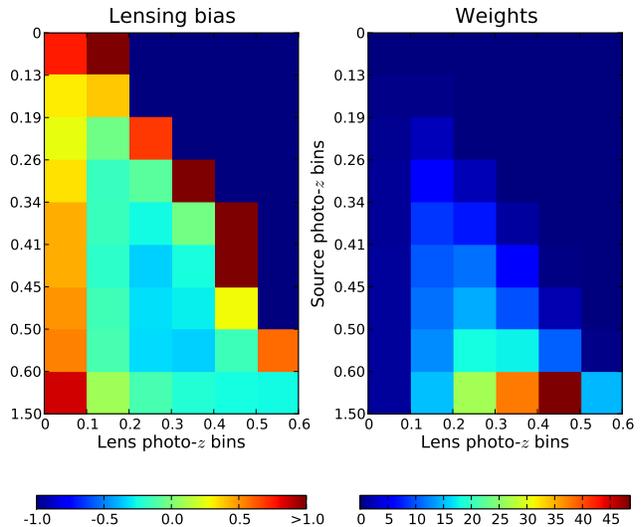}
\caption{
{\em Left:} Lensing signal bias, as a function of \photoz\ bins in both lenses and sources.
The lens bins are split into 0.1 redshift intervals, while the source bins are split into
9 equal parts.
{\em Right:} Lensing signal weights in each of the lens/source bins.
The highest source \photoz\ bins carry the most weight at any lens \photoz\ bin.
The lowest lens \photoz\ bins, as well as the bins where lens \photoz\ is near the
source \photoz, are highly biased, but carry very little weight.
\label{fig:photozbiasgrid}  
}
\end{figure}

Figure~\ref{fig:photozbiasgrid} shows the lensing bias as a function 
of lens and source \photoz\ bin.  The lens \photoz\ bins are in 0.1 intervals, while the 
source \photoz\ calibration sample was split into 9 equal parts of approximately 1000 
galaxies each.  
The \photoz\ bin edges are given in the vertical axis of the figure.
The highly-biased bins generally have low weighting;
the lowest lens \photoz\ bin is highly biased, and carries no weight
(hence also rendering the calibration unreliable), and 
similarly for bins where lens \photoz's are close to source \photoz's.
This figure clearly demonstrates that the highest source \photoz\ bin carries 
the most weight at any lens \photoz\ bin, and also has the lowest bias (except for 
the lowest lens \photoz\ bin).  This is expected, because
as the source redshift is well above the lens redshift, the weight
$\propto\Sigma_{\rm crit}^{-2}$ increases, while the 
the lensing strength $\Sigma_{\rm crit}^{-1}$ becomes less sensitive to the 
source redshift errors.

\section{Summary and conclusion}
\label{sec:conclusion}

In this paper, we have sought to address technical issues required to
use photometric redshifts for both lenses and sources in galaxy-galaxy
lensing analyses without incurring significant systematic error.

To this end, we have obtained photometric redshifts for SDSS DR8 5-band photometry,
using the template-based \photoz\ code {\tt ZEBRA}.  
Multiple readily available and self-generated \photoz\ catalogues were perused and 
compared to optimize the \photoz\ according to the key science goals of the catalogue, 
galaxy-galaxy lensing, which led us to choose a method 
with minimal \photoz\ scatter for galaxies with $\zphot>0.1$
(which dominate such analyses).  The underlying assumption is that
with a calibration sample of sufficient size to characterize the bias
and scatter, any \photoz\ biases can simply be calibrated
out, so they should not figure into the choice of \photoz\ method
(provided that they are not so pathological as to make the problem of
calibrating them out with a reasonable size calibration sample
intractable). 

To measure the \photoz\ biases and scatter, a calibration set was required.
Tests of the calibration set, and corrections for exceptional
observing conditions, were required to ensure that its \photoz\ scatters and biases
accurately represent those of the full SDSS DR8 lensing and source samples.
The calibration sample was drawn from several effectively flux-limited
spectroscopic samples (six for the shallower lens sample, five for the
deeper source sample)
consisting of $\sim$9k galaxies overlapping the SDSS footprint. 

The measured \photoz\ bias and scatter allowed estimates of their
contributions to bias in observed galaxy-galaxy lensing signals.  We considered source and
lens \photoz\ separately, with emphasis on the newer application to SDSS,
lens \photoz.
The use of \photoz's for source redshifts (in addition to lens
spectroscopic redshifts) allows for 
gravitational lensing signal calibration that is more precise than using an assumed 
redshift distribution, and allows for selection of specific source
subsamples (e.g., brighter or fainter, or with some lens-source
separation).  For a known lens redshift, we estimate that the lensing
tangential shear 
can be converted to surface mass density with an accuracy of 2 per cent. 
%
Using \photoz's for lenses will make lenses available at higher redshifts 
(out to $z\sim0.5$, limited to red galaxies) and dimmer magnitudes (down to 
$r\sim21$ or $\Mr\sim-19$, limited to blue),
an increase of factors of at least ten compared to SDSS spectroscopic
samples.  Rather than lens samples with several tens of thousands of
galaxies per stellar mass or luminosity bin as in
\cite{mandelbaum_etal:2006}, the sample sizes will be of order a few
$\times 10^{5}$, which will allow for very high $S/N$ measurements
(for which precise calibration of systematics is, therefore, a high priority).
The use of lens \photoz\ converts to surface mass density with an accuracy of 4 per cent
when the lens redshift is $0.1<\zphot<0.5$. 

Besides the lensing signal bias originating in distance relations,
biases in the observable or ``stacking quantities''  
(required for galaxy-galaxy lensing) arise because such quantities (absolute luminosity, 
rest-frame colours, stellar mass) also depend on redshift.
These biases can be studied because of the availability of best-fit galaxy SED types 
when obtaining \photoz\ from the template-based method.  
The SED's provide $k$-corrections, which in turn allow for a uniform definition of 
absolute magnitude across a span of redshifts, as well as stellar mass 
estimates. 

Because we only have a small number of heterogeneous, narrow-field spectroscopic 
survey samples to serve as the calibration sample, specific cautions were required.
For each, the full depth of the photometric survey must be covered to accurately 
represent the whole sample.  The wide separation of each of the subsamples 
compensates for the LSS in the narrow field surveys, such that the LSS can be 
averaged over to represent the (expected) smooth $\rmd N/\rmd z$ of the SDSS photometric sample.
We found that the differences in survey conditions (such as seeing and
sky noise) of the SDSS photometric sample at the locations of the spectroscopic survey area
can bias the magnitude completeness and redshift distributions
for objects near the limiting magnitude ($r\sim21.8$).  These then affected the 
lensing signal bias, where the dim, high redshift source objects have a smaller bias
and are highly weighted.  Note that our findings in this paper
supersede those in \cite{mandelbaum_etal:2008}, which (a) had a
calibration sample that was a factor of four smaller, and (b) lacked
corrections for the atypical observing conditions in SDSS at the
position of the COSMOS survey.

While the bias and scatter in \photoz\ or \photoz-derived quantities can be large, we show how 
calibrating them against the true redshifts allow us to choose a binning width 
of various lens characteristics appropriate for the uncertainties, and to estimate the 
correction for the bias.  These biases can now be used to push galaxy-galaxy lensing 
into a new regime for SDSS.

\section*{Acknowledgments}
We thank Carlos Cunha, Pascal Oesch, Alex Szalay, Istvan Csabai, Tamas Budavari, Jeff Newman,
Feng Dong, Jim Gunn, Yen-Ting Lin, Michael Blanton, and Robert Feldmann, for useful discussions.
RN was supported in part by NASA LTSA grant NNG04GC90G.  This work has been supported 
in part by World Class University grant R32-2009-000-10130-0 through the National Research
Foundation, Ministry of Education, Science and Technology of Korea.

We thank the PRIMUS team for sharing their redshift catalog, and thank Alison Coil and
John Moustakas for help with using the PRIMUS dataset.
Funding for PRIMUS has been provided by NSF grants AST-0607701, 0908246, 0908442, 0908354, and NASA grant 08-ADP08-0019.  This paper includes data gathered with the 6.5 meter Magellan Telescopes located at Las Campanas Observatory, Chile.

Funding for the DEEP2 survey has been provided by NSF grants AST95-09298, AST-0071048, AST-0071198, AST-0507428, and AST-0507483 as well as NASA LTSA grant NNG04GC89G. 
Some of the data presented herein were obtained at the W. M. Keck Observatory, which is operated as a scientific partnership among the California Institute of Technology, the University of California and the National Aeronautics and Space Administration. The Observatory was made possible by the generous financial support of the W. M. Keck Foundation. The DEEP2 team and Keck Observatory acknowledge the very significant cultural role and reverence that the summit of Mauna Kea has always had within the indigenous Hawaiian community and appreciate the opportunity to conduct observations from this mountain.

Funding for the SDSS and SDSS-II has been provided by the Alfred P. Sloan Foundation, the Participating Institutions, the National Science Foundation, the U.S. Department of Energy, the National Aeronautics and Space Administration, the Japanese Monbukagakusho, the Max Planck Society, and the Higher Education Funding Council for England. The SDSS Web Site is http://www.sdss.org/.

The SDSS is managed by the Astrophysical Research Consortium for the Participating Institutions. The Participating Institutions are the American Museum of Natural History, Astrophysical Institute Potsdam, University of Basel, University of Cambridge, Case Western Reserve University, University of Chicago, Drexel University, Fermilab, the Institute for Advanced Study, the Japan Participation Group, Johns Hopkins University, the Joint Institute for Nuclear Astrophysics, the Kavli Institute for Particle Astrophysics and Cosmology, the Korean Scientist Group, the Chinese Academy of Sciences (LAMOST), Los Alamos National Laboratory, the Max-Planck-Institute for Astronomy (MPIA), the Max-Planck-Institute for Astrophysics (MPA), New Mexico State University, Ohio State University, University of Pittsburgh, University of Portsmouth, Princeton University, the United States Naval Observatory, and the University of Washington.

\bibliographystyle{mn2e}
\bibliography{photoz}

\appendix

\section{Selection procedure for catalogue generation}
\label{S:galcut}

The following are the cuts imposed on the SDSS \texttt{Photo} pipeline
outputs 
to create the galaxy catalogues used for this work: 
\begin{itemize}
\item OBJC\_TYPE $=3$;
\item Require BINNED1 in $r$ and $i$, and overall;
\item Reject those with BLENDED but without NODEBLEND set;
\item Reject object flags: SATURATED, SATURATED\_CENTER, EDGE,
  LOCAL\_EDGE, MAYBE\_CR, MAYBE\_EGHOST, SUBTRACTED, BRIGHT,
  TOO\_LARGE, BADSKY;
\item Reject those with the following set in $r$ and $i$: CR, INTERP,
  INTERP\_CENTER;
\item Extinction-corrected $r_{\rm model}<21.8$ (source catalogue) or
  $22$ (full photometric catalogue; $<21$ for the photometric lens catalogue);
\item $r$-band extinction $A_r<0.2$;
\item {\em Source catalogue only:} Resolution factor $R_2>1/3$ in both $r$ and $i$ bands (comparing
  the image adaptive moments with the PSF adaptive moments, and
  requiring the object to be well-resolved), as in \cite{mandelbaum_etal:2005}.
\end{itemize}

Note that the cuts on the model magnitude and resolution are imposed
{\em after} eliminating duplicates by choosing the observation with
better seeing (i.e., if the object fails the magnitude cut in the
observation with better seeing, and passes it in the observation with
worse seeing, it will not be in the catalogue).

\section{Comparison with other photo-$z$'s}
\label{app:other_photozs}

\begin{figure}
\includegraphics[trim=1mm 7mm 9mm 15mm, clip, width=86mm]{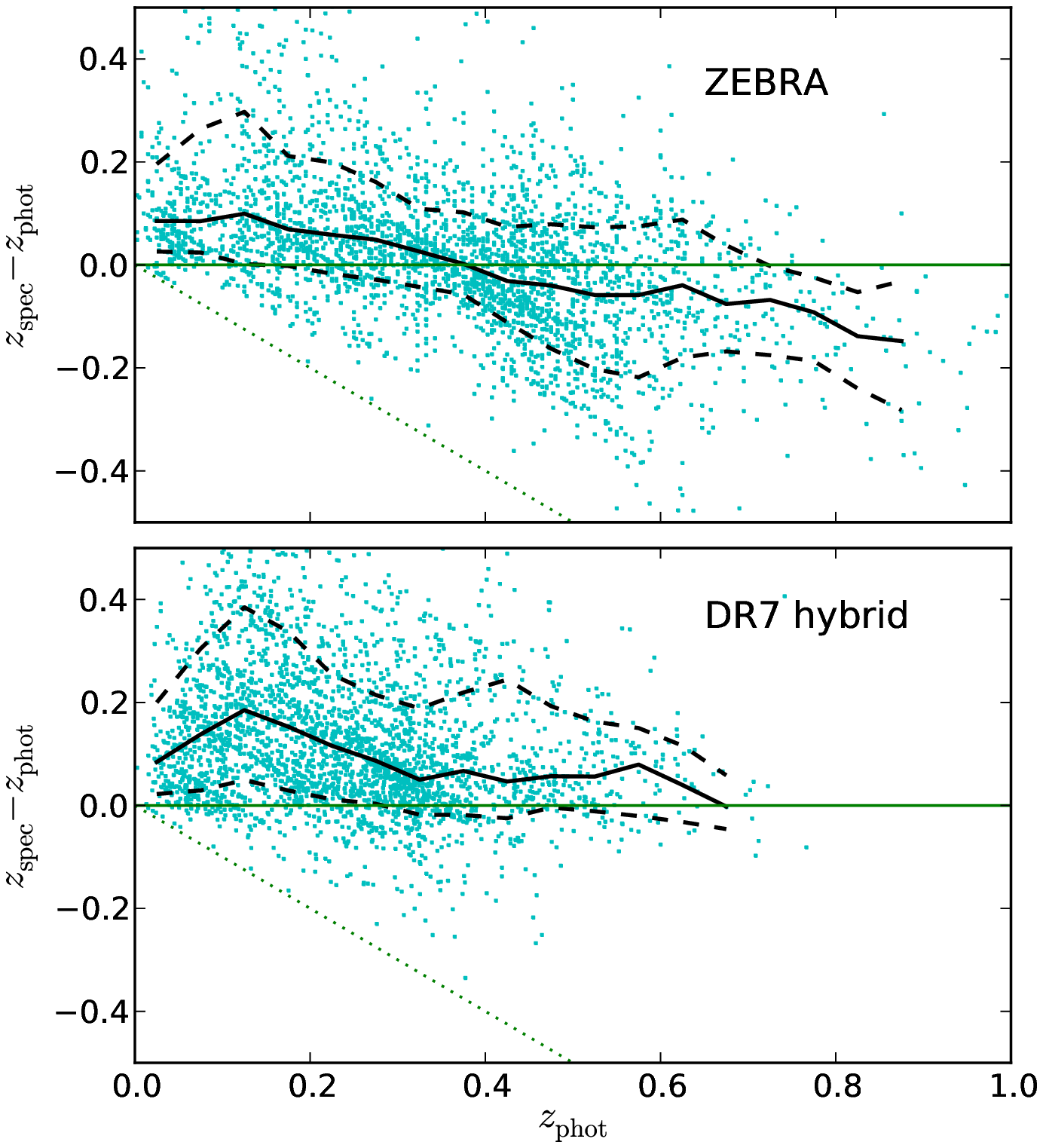}
\caption{
Comparison of the bias and scatter for the ``hybrid'' and our ZEBRA \photoz\ methods
for the source-matched catalogue.
Only 50 per cent of the points are plotted for clarity.
The limited training set in the ``hybrid'' method (the SDSS spectroscopic samples)
truncates the available \photoz's to $\zphot<0.6$. 
\label{fig:other_photozs}  
}
\end{figure}

There are several other \photoz's available for the SDSS public catalogue.
Here we illustrate the issues that made them sub-optimal for our galaxy-galaxy
lensing analysis.
The publicly available ``hybrid'' \citep{sdssdr7:2009, csabai_etal:2003}, 
``cc2'' and ``d1'' neural net methods \citep[both][]{oyaizu_etal:2008}, and the $p(z)$
redshift probability distributions \citep{cunha_etal:2009} are all essentially
training-set based methods.  
There are a few issues related to the suitability of the
training set for the photometric sample of interest; specifically, we aim for more 
high-$z$ galaxies properly classified as high-$z$, and there are simply not enough 
spectroscopic surveys available at this redshift to both train and test the SDSS \photoz's. 

The DR7 hybrid method uses a $\sim$50k training set uniformly selected to
match the colour distribution of the photometric sample, taken from the
SDSS MAIN ($r<17.7$) and LRG ($r<19$) spectroscopic sample, including an 
additional $\sim$3k blue, high-$z$ ($0.25<z<0.4$) objects to complement 
the colour distribution in the training set.  
This training set is, however, limited to bright objects, and hence to low redshifts 
($z_{\rm median}\sim0.1$), which results in the generated \photoz\ being similarly 
limited to this redshift range (see Fig.~\ref{fig:other_photozs}).
Our deeper photometric samples are demonstrably at higher redshifts.  
Their new DR8 \photoz\ sample is shown to perform better in the high-$z$ regime, based on a test 
using half the spectroscopic sample to train, and the other to test  (Csabai, private
communication).  We note, however, that their public DR8 sample is trained on the same 
spectroscopic sample as our calibration set; we require further spectroscopic redshift samples to 
independently calibrate the biases in their DR8 sample.

\begin{figure}
\includegraphics[trim=1mm 0mm 9mm 9mm, clip, width=86mm]{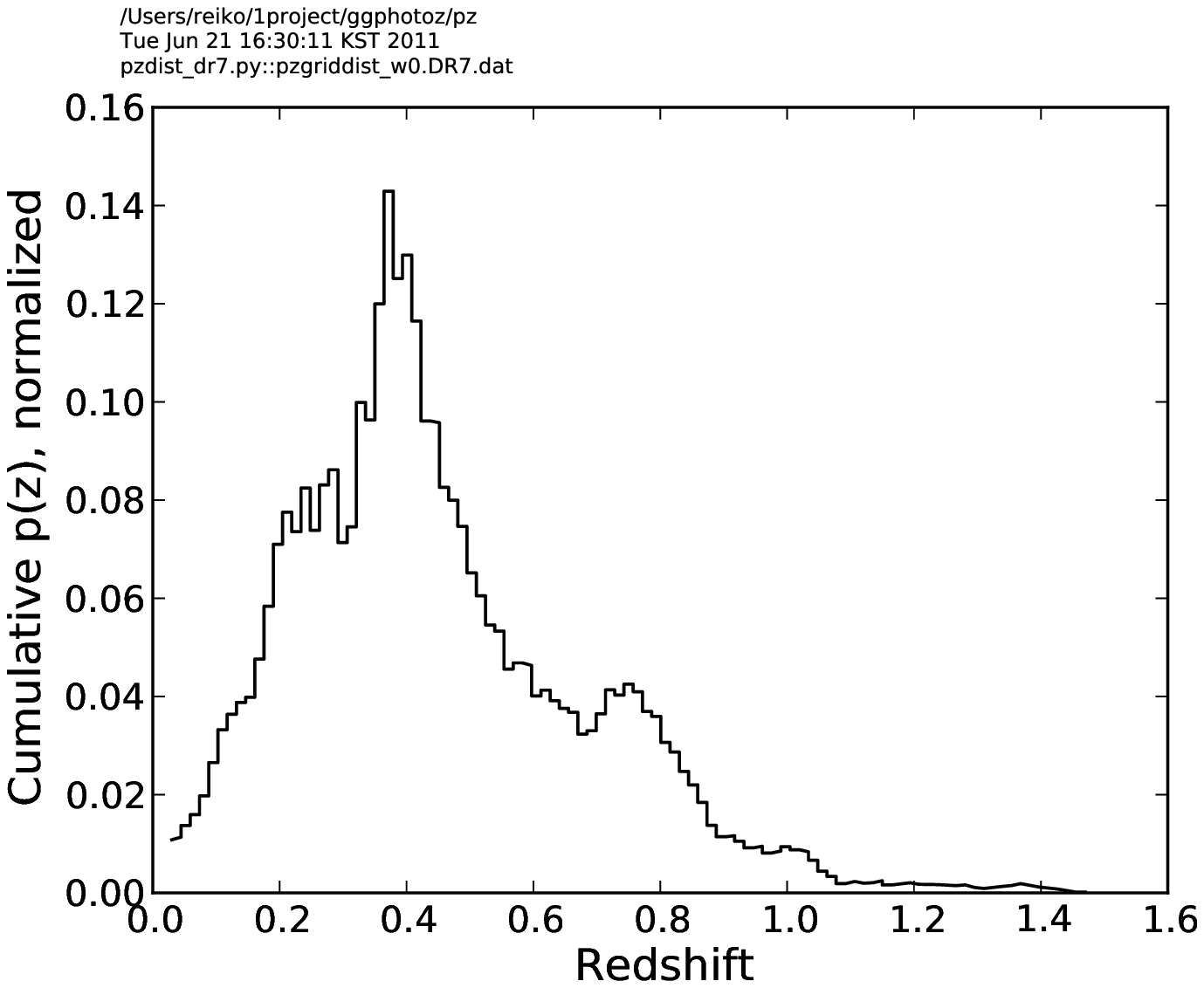}
\caption{
Cumulative $p(z)$ of all SDSS galaxies, based on DR7 $p(z)$ \citep{cunha_etal:2009}.
Although in the $p(z)$ method, the calibration samples are weighted to better represent
the whole of the SDSS photometric sample, the LSS evident in the calibration set appears
in the $p(z)$ photometric redshifts (cf. fig.~17 of \citealt{cunha_etal:2009} for the 
training set distribution).
The excess at $z\sim0.75$ is due to the non-EGS DEEP2 spectroscopic training subsample, 
which intentionally targets high-$z$ objects.
(The split at $z\sim0.3$ is due to an issue in the training set, and is absent 
in the DR8 version of the $p(z)$; Sheldon et~al. 2011 in~prep.)
\label{fig:cumulative_pz}  
}
\end{figure}

The cc2 and d1 neural net methods, as well as the $p(z)$ redshift probability distribution,
also use the SDSS MAIN (435k) and LRG ($>$80k) spectroscopic samples as the training set,
with supplements from other spectroscopic samples with deeper flux
limits and broader redshift coverage, but smaller size  ($\sim7$k).
For the $p(z)$ method, the training samples are then weighted to 
reproduce the $r$-magnitude and color distributions of the photometric sample. 
The limited number of high-$z$ regions contributing to \photoz\
estimation results in the LSS in those regions being imprinted onto
the output \photoz\ for the full photometric sample  (see Fig.~\ref{fig:cumulative_pz}).
The excess at $z\sim0.75$ is due mostly to the non-EGS DEEP2 spectroscopic training subsample,
because they select high-$z$ ($z\gtrsim0.7$) galaxies to target.  Since the DEEP2 targeting
were based on non-SDSS bands, the magnitude-based reweighting scheme is not complete, 
causing the excess structure.
Additionally, any LSS present in the training set reflects directly into the $p(z)$
(Sheldon et~al\@. 2011, in~prep.).  
In general, if the sample we use to test our calibration is also in the training set
(which is the case here), then we expect excessively optimistic results.
There are simply not enough spectroscopic samples at high-$z$ available to 
split between testing and training.  

For the lens sample, in addition to aiming for \photoz\ accuracy,
we require $k$-correction for our galaxy-galaxy lensing analysis,
to stack lenses according to $k$-corrected galaxy observables
such as luminosity and colour. Such information is not available with
the d1 and cc2 methods, despite having the least bias and scatter for the lens
sample.  Although training-set based method can, in principle, perform spectral 
classifications \citep{firth_etal:2003, collister_lahav:2004}, the estimated SED
it is not available with these \photoz's, and hence $k$-correction is not available. 
The hybrid method includes an SED estimator which enables a $k$-correction, 
but their \zphot\ scatter at low \photoz\ is higher than with the ZEBRA method.

We briefly compare our \photoz\ to the MegaZ-LRG sample by 
\citet{collister_etal:2007} with a scatter of $\sigma_{\Delta z/(1+z)} = 0.045$.  
We impose the MegaZ-LRG color and magnitude cuts to our calibration sample,
and find that ZEBRA classifies the calibration sample into Ell and Sbc types with 259 and
102 objects, with scatter of $\sigma_{\Delta z/(1+z)} = 0.04$ and 0.07, respectively.
For the combined sample, the scatter is 0.05.
While the scatter in our calibration sample is not quite as small as that from 
MegaZ-LRG, our cuts do not include the fiber magnitude cut or the magnitude-based 
star-galaxy separation criteria cuts imposed on the MegaZ-LRG sample, which 
presumably would decrease the number of outliers we see in our \photoz\ error.
Additionally, we note that our need for $k$-corrections to define luminosities 
prevents us from using photo-z estimators without any template information, 
such as the neural net method of \citet{collister_etal:2007}.

\end{document}